\documentclass[oneside,american,reqno]{elsart}
\usepackage{glossary}
\usepackage{amssymb}
\usepackage{pslatex}
\usepackage{graphics}
\usepackage[T1]{fontenc}
\usepackage[latin1]{inputenc}
\usepackage{amsmath}
\usepackage{algorithmic} 
\newcommand{\singlespace}{\renewcommand{\baselinestretch}{1.0}\small \normalsize}
\newcommand{\komment}[1]{}
\newenvironment{proof}{\begin{pf}}{\qed\end{pf}}
\makeatletter
 \usepackage{verbatim}
\usepackage{pslatex}
\newcommand{\daniel}[1]{} 
\newcommand{\owen}[1]{}
\newcommand{\chainsaw}[1]{#1}

\newcommand{\verbose}[1]{}

\newcommand{\chooz}[2] { #1 \choose #2 }

\newcommand{\R}{{\sf R\hspace*{-0.9ex}\rule{0.15ex}%
       {1.3ex}\hspace*{0.9ex}}}

\usepackage{ifpdf}
\ifpdf
\usepackage{thumbpdf}
\usepackage[pdftex]{hyperref}
\fi

\makeatother
\makeglossary
\begin{document}

\begin{frontmatter}
\title{Attribute Value Reordering For Efficient Hybrid OLAP}
\date{\today{}}
\author[unbsj]{Owen Kaser\corauthref{cor}}
\corauth[cor]{Corresponding author.}
\address[unbsj]{ Dept. of Computer Science and Applied Statistics\\
	 U. of New Brunswick,
	 Saint John, NB Canada}

\author[nrc]{Daniel Lemire}
\address[nrc]{Université du Québec à Montréal\\
         Montréal, QC Canada
}
\begin{abstract}
The normalization of a data cube is the
ordering of the attribute values.
For large multidimensional arrays where dense and sparse chunks are stored
differently, proper normalization can lead to improved storage efficiency.
We show that
it is NP-hard to compute an optimal normalization even for $1\times3$
chunks, although we find an exact algorithm for $1\times2$ chunks.
When dimensions are nearly statistically independent, we show
that dimension-wise attribute
frequency sorting is an optimal normalization and takes time $O(d n\log(n))$ for data
cubes of size $n^d$.
When dimensions are not independent, we propose and evaluate a
several heuristics.
The hybrid OLAP (HOLAP) storage mechanism
is already 19\%--30\% more efficient than ROLAP, but
normalization can improve it further by 9\%--13\%
for a total gain of 29\%--44\% over ROLAP.
\end{abstract}
\begin{keyword}
Data Cubes \sep Multidimensional Binary Arrays \sep MOLAP \sep Normalization \sep Chunking
\end{keyword}
\end{frontmatter}

\renewcommand{\thefootnote}{}
\footnotetext{This is an expanded version of our earlier paper~\cite{kase:reorderDOLAP}.}
\renewcommand{\thefootnote}{\arabic{footnote}}

\section{Introduction}

On-line Analytical Processing (OLAP) is a database acceleration technique
used for deductive analysis~\cite{goil:thesis}.
The main objective of OLAP is to have constant-time or near
constant-time answers for many typical queries. For example, in a database
containing salesmen's performance data, one may want to compute on-line
the amount of sales done in Ontario for the last 10 days, including
only salesmen who have 2 or more years of experience.
Using a relational database containing sales information, such a
computation may be expensive. Using OLAP, however,
the computation is typically done on-line. To achieve such acceleration
one can create a \emph{cube} of data,  
 a map from all attribute values to a given
measure. In the example above, one could map tuples containing days,
experience of the salesmen, and locations to the corresponding amount
of sales. 

We distinguish two types of OLAP engines: Relational OLAP (ROLAP) and
Multidimensional OLAP (MOLAP). In ROLAP, the data is itself stored in a
relational database whereas with MOLAP, a large multidimensional array
is built with the data. In MOLAP, an important step in building a data cube
is choosing a \emph{normalization}, which is a mapping from attribute values 
to the integers used to index the array.
One difficulty with MOLAP is that the array is often sparse.  For
example, not all tuples (day, experience,
location) would match sales. Because of this sparseness, ROLAP uses far 
less storage.  Additionally, there are compression
algorithms to further decrease ROLAP storage 
requirements~\cite{dehn:CGM-OLAP-datamining,ng:compression-statistical,sism:dwarf}.
On the other hand, MOLAP
can be much faster, especially if subsets of the data cube are 
dense~\cite{zhao:arraybased}. Many vendors such as Speedware, Hyperion, IBM, and Microsoft
are thus using Hybrid OLAP (HOLAP), storing dense regions of the cube
using MOLAP and storing the rest using a ROLAP approach. \verbose{In other
words, a convenient and efficient representation of a sparse data cube
is achieved when the dense regions are stored as multidimensional
arrays and the sparse remainder is represented as a list of values as
in a relational database.}

While various efficient heuristics exist to find dense sub-cubes
in data cubes~\cite{cheu:DROLAP,cheu:towards-DROLAP,kase:compress-tr},
\verbose{one problem we still face is that} the dense sub-cubes are 
normalization-dependent.\verbose{,
so that the same data with attribute values ordered differently may
have completely different dense sub-cubes and may be stored significantly
more efficiently.} A related problem with MOLAP or HOLAP
is that the attribute values may not have a canonical ordering, so
that the exact representation chosen for the cube is arbitrary. In
the salesmen example, imagine that {}``location'' can have
the values {}``Ottawa,'' {}``Toronto,'' {}``Montreal,'' {}``Halifax,''
and {}``Vancouver.'' How do we order these cities: by population,
by latitude, by longitude, or alphabetically? Consider the example given
in Table~\ref{cap:Two-tables-representing}: it is obvious that HOLAP
performance will depend on the normalization of the data cube.
\verbose{We also believe that} A storage-efficient normalization may \verbose{often}
lead to better query performance.

\begin{table}
\caption[Two tables representing the volume of sales.]{\label{cap:Two-tables-representing}Two tables representing the volume
of sales for a given day by the experience level of the salesmen.
Given that three cities only have experienced salesmen, some orderings
(left)
will lend themselves better to efficient storage (HOLAP) than others
(right).}
\begin{singlespace}\footnotesize
\begin{center}\begin{tabular}{|c|c|c|c|}
\hline
&
<1~yrs&
1--2~yrs&
>2~yrs\tabularnewline
\hline
\hline 
Ottawa&
&
&
\$732\tabularnewline
\hline
Toronto&
&
&
\$643\tabularnewline
\hline 
Montreal&
&
&
\$450\tabularnewline
\hline
Halifax&
\$43&
\$54&
\tabularnewline
\hline 
Vancouver&
\$76&
\$12&
\tabularnewline
\hline
\end{tabular}~\begin{tabular}{|c|c|c|c|}
\hline 
&
<1~yrs&
1--2~yrs&
>2~yrs\tabularnewline
\hline
\hline 
Halifax&
\$43&
\$54&
\tabularnewline
\hline 
Montreal&
&
&
\$450\tabularnewline
\hline 
Ottawa&
&
&
\$732\tabularnewline
\hline 
Vancouver&
\$76&
\$12&
\tabularnewline
\hline 
Toronto&
&
&
\$643\tabularnewline
\hline
\end{tabular}\end{center}
\end{singlespace}

\end{table}

One may object that normalization only applies when attribute values
are not regularly sampled numbers. One argument against normalization
of numerical attribute values is that storing an index map from these
values to the actual index in the cube amounts to extra storage.
This extra storage is not important. Indeed, consider a data
cube with $n$ \glossary{name={$n$},description={number of attribute values per dimension}} attribute values per dimension
and $d$ \glossary{name={$d$},description={number of dimensions}}
 dimensions: we say such a cube is
\emph{regular} or \emph{ $n$-regular}. 
\glossary{name={$n$-regular},description={cube with exactly $n$ attribute values per dimension}}
The most naive way to store
such a map is for each possible attribute value to store a new index
as an integer from $1$ to $n$. Assuming that indices are stored using
$\log n$~bits, this means that $n\log n$ bits are required.  However,
array-based storage of a regular data cube uses $\Theta(n^{d})$
bits. In other words, unless $d=1$, normalization is not a noticeable
burden and all dimensions can be normalized. \verbose{The case $d=2$ is similar
to image compression, where reordering pixels is a widely used
technique~\cite{netr:fax-ordering}. For high dimensional data cubes,
the possible gains are substantial.}

\chainsaw{ Normalization may degrade performance if
attribute values often used together are stored in physically
different areas thus requiring extra IO operations. When attribute
values have hierarchies, it might even be desirable to restrict the possible
reorderings. However, in
itself, changing the normalization does not degrade the
performance of a data cube, unlike many compression algorithms.  While
automatically finding the optimal normalization may be difficult when
first building the data cube, the system can run an optimization
routine after the data cube has been built, possibly as a background
task.  }

\subsection{Contributions and Organization}
The contributions of this paper include a detailed look at
the mathematical foundations of normalization, including
notation for the remainder of the paper and future work
on normalization of block-coded data cubes
(Sections~\ref{section:block-coded}~and~\ref{section:math-prelims}).
In particular, Section~\ref{section:math-prelims} includes a theorem
showing that determining whether two data cubes are equivalent
for the normalization problem is \textsc{Graph Isomorphism}-complete.
Section~\ref{section:generalcomplexity} \verbose{then} considers the
computational complexity of normalization.
If data cubes
are stored in tiny (size-2) blocks, an exact algorithm can
compute the best normalization, whereas for larger blocks,
it is conjectured that the problem is NP-hard.  As evidence,
we show that the case of size-3 blocks is NP-hard.
Establishing that even trivial cases are
NP-hard helps justify use of heuristics.
Moreover, the optimal algorithm used for tiny blocks 
leads us to the  Iterated Matching (IM) heuristic presented later.
An important class of ``slice-sorting'' normalizations is
investigated in Section~\ref{section:slicesort}.  \verbose{These
normalizations can be efficiently calculated, but the quality of
their solutions is sometimes poor.}  Using a notion of
statistical independence, a major contribution
(Theorem~\ref{theorem:bigbadtheorem}) is an easily computed
approximation bound for a heuristic called ``Frequency Sort,''
which we show to be the best choice among our heuristics when the cube dimensions are nearly statistically independent.
Section~\ref{section:heuristics} discusses additional
heuristics that could be used when the dimensions of the cube are not
sufficiently independent.  In Section~\ref{section:experiments},
experimental results compare the performance of heuristics
on a variety of synthetic and ``real-world'' data sets.
The paper concludes with Section~\ref{section:conclusion}.
A glossary is
provided at the end of the paper.

\section{Block-Coded Data Cubes}
\label{section:block-coded}

In what follows, $d$ is the number of dimensions (or attributes) of
the data cube $C$ and $n_i \mbox{, for\ } 1 \leq i \leq d$, is the
number of attribute values for dimension $i$.  Thus, $C$ has size
$n_{1}\times\ldots\times n_{d}$.  To be precise, we distinguish
between the \emph{cells} and the \emph{indices} of a data cube.
``Cell'' is a logical concept and each cell corresponds uniquely to a
combination of values $(v_1,v_2,\ldots,v_d)$,  with one value $v_{i}$
for each attribute $i$.  
In
Table~\ref{cap:Two-tables-representing}, one of the 15 cells
corresponds to (Montreal, 1--2~yrs). \emph{Allocated} cells, such as
(Vancouver, 1--2~yrs), store measure values, in contrast to unallocated
cells such as (Montreal, 1--2~yrs).  From now on, we shall assume that
some initial normalization has been applied to the cube and that
attribute $i$'s values are $\{ 1, 2, \ldots n_i \}$.  ``Index'' is a
physical concept and each $d$-tuple of indices specifies a storage
location within a cube.  At this location there is a cell, allocated
or otherwise.  \emph{(Re-) normalization changes neither the cells nor
the indices of the cube; (Re-)normalization changes the assignment of
cells to indices.}

We use $\#C$ 
\glossary{name={$\#C$},description={number of allocated cells in a sparse cube}} 
to denote the number of allocated cells in 
cube $C$.  Furthermore, we say that $C$ has \emph{density} 
$\rho=\frac{\#C}{n_{1}\times\ldots\times n_{d}}$.
\glossary{name=$\rho$,description={density of a cube: number of allocated cells divided by total number of cells}}
\chainsaw{
While we can optimize storage requirements and speed up
queries by providing approximate 
answers~\cite{barb:loglinear,vitter99:approximate,ried:pcube00},
we focus on exact methods in this paper, and so}
we seek an efficient storage mechanism to store all $\#C$ allocated cells.

There are many ways to store data cubes using different coding for
dense regions than for sparse ones. For example, in one
paper~\cite{kase:compress-tr} a single dense 
sub-cube (chunk) with $d$ dimensions
is found and the remainder is considered sparse. \verbose{It is also 
possible to use a broader class of regions.}

\verbose{To determine the best storage strategy for data cubes,
we have considered image compression, where simple blocks are used in
many image compression formats including
JPEG~\cite{donoho98:data}.
While some attempt to improve the
current formats by dividing images into arbitrarily shaped regions
 through adaptive algorithms~\cite{lepennc-bandelets}, others provide
evidence that non-adaptive algorithms 
suffice~\cite{candes:surprisingly}.
It is not clear
\emph{a priori} that more complex shapes lead to more efficient storage.}

We follow earlier work~\cite{goil:thesis,sara:efficient-storage} 
and store the data cube in \emph{blocks}\footnote { Many authors use the
term ``chunks'' with different meanings.  }, which are disjoint $d$-dimensional sub-cubes
covering the entire data cube. We consider blocks of constant size
$m_{1}\times\ldots\times m_{d}$; 
\glossary{name={$m_i$},description={block size in dimension $i$}} 
thus, there are $\lceil
\frac{n_{1}}{m_{1}}\rceil
\times\ldots\times\lceil\frac{n_{d}}{m_{d}}\rceil$ blocks.  For
simplicity, we usually assume that $m_{k}$ divides $n_{k}$ for all $k
\in \lbrace 1,\ldots,d \rbrace$.  Each block can then be stored in an
optimized way depending, for example, on its density.
We consider only two widely used coding
schemes for data cubes, corresponding respectively to simple ROLAP and
simple MOLAP.
That is, either we represent the block as a list of
tuples, one for each allocated cell in the block, or else we code the
block as an array. For both extreme cases, a very dense or a very
sparse block, MOLAP and ROLAP are respectively \emph{efficient}.
\verbose{Another argument for block-coded data cubes is that many efficient
buffering schemes for OLAP range queries rely themselves on block 
coding~\cite{geffner:rps,lemi:CASCON}.}
More aggressive compression
 is possible~\cite{li:CompressedDataWarehouses}, but as long as we use block-based storage, normalization is
a factor.

Assuming that a data cube is stored using block encoding, we need to
estimate the storage cost. A simplistic model is given as follows.
The cost of storing a single cell sparsely, as a tuple containing the
position of the value in the block as $d$ attribute values (cost proportional
to $d$) and the measure value itself (cost of 1), is assumed to be $1 + \alpha d$,
\glossary{name={$\alpha$},description={parameter in our sparse storage cost model which is $1 + \alpha d$ per allocated cell, we set $\alpha = 1/2$}}
where parameter $\alpha$ can be adjusted to account for size differences between 
measure values and attribute values. Setting $\alpha$ small would favor 
sparse encoding (ROLAP) whereas setting $\alpha$ large would favor dense encoding (MOLAP).
For example, while we might store 32-bit measure values, the number of values
per attribute in a given block is likely less than $2^{16}$. 
This motivates setting $\alpha = 1/2$ in later experiments and the
remainder of the section.  Thus, densely storing a block with $D$ 
\glossary{name=$D$,description={number of allocated cells in a block}} allocated cells
costs $M =m_{1}\times\ldots\times m_{d}$, 
\glossary{name={$M$},description={total number of cells  per block}} but storing it sparsely
costs $(d/2+1)D$.  

It is more economical to store a block densely if $(d/2+1)D>M$, that
is, if $\frac{D}{m_{1}\times\ldots\times m_{d}}>\frac{1}{d/2+1}$.  This block coding is least efficient when a data cube
has uniform density $\rho$ over all blocks. In such cases, it has a
sparse storage cost of $d/2+1$ per allocated cell if $\rho \leq
\frac{1}{d/2+1}$ or a dense storage cost of $1/\rho$ per allocated cell
if $\rho > \frac{1}{d/2+1}$.  Given a data cube $C$, $H(C)$ denotes
\glossary{name={$H(C)$},description={HOLAP storage cost for cube $C$}}
its storage cost.  We have $\#C\leq H(C)\leq n_{1}\times\ldots\times
n_{d}$.  Thus, we measure the cost per allocated cell $E(C)$ as
\glossary{name={$E(C)$},description={cost per allocated cell in cube $C$ or $\frac{H(C)}{\#C}$}}
$\frac{H(C)}{\#C}$ with the convention that if $\#C=0$, then $E(C)=1$.
The cost per allocated cell is bounded by 1 and $d/2+1$: $1\leq E(C)\leq d/2+1$.  
A weakness of the model is that
it ignores obvious storage overheads proportional to the number of
blocks, $\frac{n_{1}}{m_{1}}\times\ldots\times\frac{n_{d}}{m_{d}}$.
However, as long as the number of blocks remains constant,
it is reasonable to assume that the overhead is constant.  
Such is the case when we consider the same
data cube under different normalizations using fixed block dimensions.

\section{Mathematical Preliminaries}
\label{section:math-prelims}
Now that we have defined a simple HOLAP model, we review
two of the most important concepts in this paper: slices
and normalizations. Whereas a slice amounts to fixing one
of the attributes, a normalization can be viewed as a tuple
of permutations.

\subsection{Slices}

\begin{figure}
\begin{center}\resizebox{6cm}{!}{\includegraphics{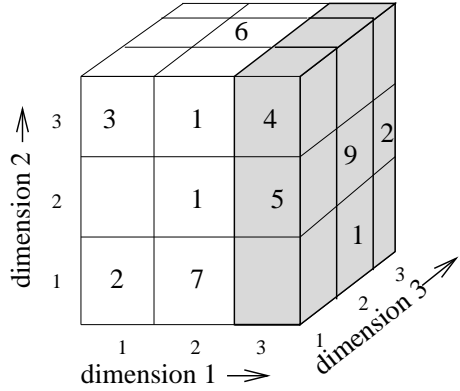}}\end{center}

\caption{\label{slicefig}
A $3 \times 3 \times 3$ cube $C$ with the slice
$C^{1}_3$ shaded.} 
\end{figure}

Consider an $n$-regular
$d$-dimensional cube $C$
and let $C_{i_{1},\ldots,i_{d}}$ denote the cell stored at indices
$(i_{1},\ldots,i_{d})\in\left\{ 1,\ldots,n\right\} ^{d}$.
\glossary{name={$C_{i_{1},\ldots,i_{d}}$},description={cell of cube $C$ at location ($i_{1},\ldots,i_{d}$)}}
Thus, $C$ has size $n^{d}$. 
The \emph{slice} $C_{v}^{j}$ of $C$, for index $v$ of dimension $j$ 
\glossary{name={$C_{v}^{j}$},description={slice of cube $C$, for index $v$ of dimension $j$}} 
($1 \leq j \leq d$  and $1 \leq v \leq n$) is a $d-1$~-~dimensional cube
formed as ${C_{v}^{j}}_{i_1,\ldots,i_{j-1},i_{j+1},\ldots,i_d} =
C_{i_1,\ldots,i_{j-1},v,i_{j+1},\ldots,i_d}$ (See Figure~\ref{slicefig}).
\verbose{See Figure~\ref{slicefig}
for an example, and observe that normalization does not affect the
collection of cells in a slice--- only the name of the slice and the cells'
indices within the slice.}

For the normalization task, \verbose{the actual data values are unimportant and}
we simply need know which indices contain allocated cells.
Hence we often view a slice as
a $d-1$~-~dimensional Boolean array \verbose{and we denote the corresponding slice by}
$\widehat{C}_{v}^{j}$.
\glossary{name={$\widehat{C}_{v}^{j}$},description={allocation map for slice $C_{v}^{j}$: Boolean array indicating whether the cell is allocated or not, see also ``allocation cube''}}
\verbose{We can also view a slice as a vector of length
$n^{d-1}$, containing either measure values or Booleans, depending on our
requirements.}
For example, in Figure~\ref{slicefig}, we might write (linearly)
$C^{1}_3=[0,1,0,5,9,2,4,0,0]$ and $\widehat{C}^{1}_3=[0,1,0,1,1,1,1,0,0]$, if we represent
non-allocated cells by zeros.
 Let $\#\widehat{C}_{v}^{j}$ denote the number of
allocated cells in slice $C_{v}^{j}$.

\subsection{Normalizations and Permutations}

Given a list of $n$ items, there
are $n!$ distinct possible permutations noted $\Gamma_{n}$ (the \emph{Symmetry Group}). 
\glossary{name={$\Gamma_{n}$},description={set of all permutations of a list of $n$ items}}
\glossary{name=$\gamma$,description={a permutation}}
If $\gamma \in \Gamma_{n}$ permutes $i$ to $j$, we write $\gamma (i) = j$.
 The identity permutation is denoted $\iota$.
\glossary{name=$\iota$,description={the identity permutation}}
In contrast to previous work on database 
compression (e.g., \cite{ng:compression-statistical}), with our
HOLAP model there is no performance advantage from
permuting the order of the dimensions themselves. (Blocking treats
all dimensions symmetrically.) 
Instead, we focus on normalizations, which affect 
the order of each attribute's values.
A normalization $\pi$ of a data cube $C$ is  
\glossary{name={normalization},description={a permutation of the order of the attribute values of each dimension in a HOLAP cube, typically to improve storage efficiency}} 
\glossary{name={$\pi$},description={a normalization (a $d$-tuple of permutations)}}
a $d$-tuple $(\gamma_{1},\ldots,\gamma_{d})$
of permutations where $\gamma_{i}\in\Gamma_{n}$ for $i=1,\ldots,d$, and
the normalized data cube $\pi(C)$ is  
\glossary{name={$\pi(C)$},description={cube $C$ after normalization by $\pi$}}
$\pi(C)_{i_{1},\ldots,i_{d}}=C_{\gamma_{1}(i_{1}),\ldots,\gamma_{d}(i_{d})}$
for all $(i_{1},\ldots,i_{d})\in\left\{ 1,\ldots,n\right\} ^{d}$. Recall
that permutations, and thus normalizations, are not commutative.
However, normalizations are always invertible, and there
are $(n!)^{d}$  normalizations for an $n$-regular data cube.
%
%
 The identity normalization is denoted $I=(\iota,\ldots,\iota)$;
whether $I$ denotes the identity normalization or the identity matrix
\glossary{name={$I$},description={identity matrix or normalization}}
will be clear from the context. Similarly $0$ may denote the
zero matrix.

Given a data cube $C$, we define its corresponding \textit{allocation cube} $A$
as a cube with the same dimensions, containing 0's and 1's depending on whether
or not the cell is allocated.
Two data cubes $C$ and $C'$, and their corresponding
allocation cubes $A$ and $A'$, are equivalent ($C\sim C'$)
\glossary{name={$\sim$},description={cube equivalence up to normalization}}
if there is a normalization $\pi$ such that $\pi(A)=A'$.

The cardinality
of an equivalence class is the number of distinct data cubes $C$ in
this class. The maximum cardinality is  $(n!)^{d}$ and there
are such equivalence classes: consider the equivalence class generated
by a {}``triangular'' data cube
$C_{i_{1},\ldots,i_{d}}=1$ if $i_{1}\leq i_{2} \leq \ldots \leq i_{d}$
and $0$ otherwise. Indeed, suppose that $C_{\gamma_1 (i_{1}),\ldots,\gamma_d(i_{d})}=
C_{\gamma'_1 (i_{1}),\ldots,\gamma'_d(i_{d})}$ for all $i_1,\ldots,i_d$, then 
$\gamma_1(i_{1})\leq \gamma_2(i_{2}) \leq \ldots \leq \gamma_d(i_{d})$
if and only if 
$\gamma'_1(i_{1})\leq \gamma'_2(i_{2}) \leq \ldots \leq \gamma'_d(i_{d})$
which implies that $\gamma_i= \gamma'_i$ for $i\in \{1,\ldots,d\}$. To see this,
consider the 2-d case where $\gamma_1(i_{1})\leq \gamma_2(i_{2})$ if and only if
$\gamma'_1(i_{1})\leq \gamma'_2(i_{2})$. In this case the result
follows from the following technical proposition. 
For more than two dimensions,
the proposition can be applied to any \emph{pair} 
of dimensions.

\begin{prop}
Consider any $\gamma_1, \gamma_2, \gamma'_1, \gamma'_2 \in \Gamma_n$ satisfying
$\gamma_1 (i) \leq \gamma_2(j) \Leftrightarrow \gamma'_1 (i) \leq \gamma'_2(j)$ for
all $1 \leq i,j \leq n$. Then $\gamma_1=\gamma'_1$ and 
$\gamma_2=\gamma'_2$.
\end{prop}
\begin{proof}
Fix $i$, then let $k$ be the number of $j$ values such that $\gamma_2(j) \geq \gamma_1(i)$.
We have that $\gamma_1(i) = n-k+1$ because it is the only element of $\{1,\ldots, n\}$
having exactly $k$ values larger or equal to it. Because $\gamma_1 (i) \leq \gamma_2(j) \Leftrightarrow \gamma'_1 (i) \leq \gamma'_2(j)$, $\gamma'_1(i) = n-k+1$ and hence
$\gamma'_1 = \gamma_1$. Similarly, fix $j$ and count $i$ values to prove that $\gamma'_2 = \gamma_2$.
\end{proof}

However,  there are singleton equivalence
classes, since some cubes are invariant under normalization: consider
a null data cube 
$C_{i_{1},\ldots,i_{d}}=0$
for all $(i_{1},\ldots,i_{d})\in\left\{ 1,\ldots,n\right\} ^{d}$.

To count the cardinality of a class of data cubes, it suffices
to know how many slices
$C_{v}^{j}$ of data cube $C$ are identical, so that we can take into
account the invariance under permutations.  Considering all $n$ slices
in dimension $r$, we can count the number of distinct slices
$d_{r}$ and number of copies
$n_{r,1},\ldots,n_{r,d_{r}}$ of each.
Then, the number of distinct
permutations in dimension $r$ is
$\frac{n!}{n_{r,1}!\times\ldots,\times n_{r,d_{r}}!}$ and
the cardinality of a given equivalence class is
$\prod_{r=1}^d\left( \frac{n!}{n_{r,1}!\times\ldots,\times n_{r,d_{r}}!}\right).$
For example, the equivalence class generated by \begin{singlespace}
$C=\begin{bmatrix}
0 & 1\\
0 & 1\end{bmatrix}$\end{singlespace}
has a cardinality of 2, despite having 4 possible normalizations.

To study the computational complexity of determining 
cube similarity, we define two decision problems.
The problem \textsc{Cube Similarity} has $C$ and $C'$ as input and asks
whether $C\sim C'$.  Problem \textsc{Cube Similarity (2-d)} restricts $C$
and $C'$ to two-dimensional cubes.
Intuitively, \textsc{Cube Similarity}  asks whether two data cubes
offer the same problem from a normalization-efficiency viewpoint.
The next theorem concerns the computational complexity of \textsc{Cube
Similarity (2-d)}, but we need the following lemma first.
Recall that $(\gamma_1,\gamma_2)$ is the normalization with the permutation $\gamma_1$
along dimension~1 and $\gamma_2$ along dimension~2 whereas  $(\gamma_1,\gamma_2)(I)$
is the renormalized cube.

\begin{lem}\label{imatrix-norm}
Consider the $n\times n$ matrix  $I' = (\gamma_1,\gamma_2)(I)$. Then $I' = I \iff \gamma_1 = \gamma_2$.
\end{lem}

We can now state Theorem~\ref{graphiso},
which shows that determining cube similarity is
\textsc{Graph Isomorphism}-complete~\cite{john:catalog-chapter}.
A problem $\Pi$
belongs to this complexity class when both
\begin{itemize}
\item
$\Pi$ has a polynomial-time reduction to \textsc{Graph Isomorphism}, and
\item
 \textsc{Graph Isomorphism} has a polynomial-time reduction to $\Pi$.
\end{itemize}

\textsc{Graph Isomorphism}-complete problems
are unlikely to be NP-complete~\cite{vanl:graph-algo-chapter}, yet there is no
known polynomial-time algorithm for any problem in the class.
This complexity class has been extensively studied.

\begin{thm} \label{graphiso}
\textsc{Cube Similarity (2-d)} is \textsc{Graph Isomorphism}-complete.
\end{thm}
\begin{proof}
It is enough to consider two-dimensional allocation cubes as 0-1
matrices.  The connection to graphs comes via adjacency matrices.

To show that \textsc{Cube Similarity (2-d)} is \textsc{graph
isomorphism}-complete, we show two polynomial-time many-to-one
reductions: the first transforms an instance of \textsc{Graph
Isomorphism} to an instance of \textsc{Cube Similarity (2-d)}.

The second reduction transforms an instance of
\textsc{Cube Similarity (2-d)} to an instance of \textsc{Graph
Isomorphism}.

The graph-isomorphism problem
is equivalent to a \textit{re-normalization} problem of the adjacency matrices.
Indeed, consider two graphs $G_1$ and $G_2$ and their adjacency matrices
$M_1$ and $M_2$. The two graphs are isomorphic if and only if there is
a permutation $\gamma $ so that  $(\gamma,\gamma)(M_1) = M_2$.
We can assume without loss of generality that all rows and columns of the
adjacency matrices have at least one non-zero value, since we can count
and remove disconnected vertices in time
proportional to the size of the graph.

We have to show that the problem of deciding whether $\gamma$
satisfies $(\gamma,\gamma)(M_1) = M_2$ can be rewritten as a data
cube equivalence problem.  It turns out to be possible by extending
the matrices $M_1$ and $M_2$.  Let $I$ be the identity matrix, and
consider two allocation cubes (matrices) $A_1$ and $A_2$ and their extensions
\begin{singlespace}
$\hat A_1 =  \begin{bmatrix} A_1 & I & I \\
					 I & I & 0 \\
                                         I & 0 & 0 \end{bmatrix}$
and
$\hat A_2 = \begin{bmatrix} A_2 & I & I \\
					 I & I & 0 \\
                                         I & 0 & 0 \end{bmatrix}.$\end{singlespace}

Consider a normalization $\pi$ satisfying $\pi(\hat A_1 ) = \hat A_2 $
for matrices $A_1, A_2$ having at least one non-zero value for each
column and each row.  We claim that such a $\pi$ must be of the form
$\pi = ( \gamma_1, \gamma_2 )$ where $\gamma_1 = \gamma_2$. By the number of non-zero values in each row and
column, we see that rows cannot be permuted across the three blocks of
rows because the first one has at least 3 allocated values, the second
one exactly 2 and the last one exactly 1. The same reasoning applies
to columns. In other words, if $x \in [j,jn]$, then $\gamma_i(x) \in
[j,jn]$ for $j=1,2,3$ and $i=1,2$.

Let $\gamma_i|j$ denote the permutation $\gamma$ restricted to block $j$
where $j=1,2,3$. Define $\gamma^{j}_i=\gamma_i|j - jn$ for $j=1,2,3$
and $i=1,2$. By Lemma~\ref{imatrix-norm}, each sub-block consisting of
an identity leads to an equality between two permutations.
From the two identity matrices in the top sub-blocks, for example,  we have that
$\gamma^{1}_1=\gamma^{2}_2$ and $\gamma^{1}_1=\gamma^{3}_2$. From the
middle sub-blocks, we have $\gamma^{2}_1=\gamma^{1}_2$ and
$\gamma^{2}_1=\gamma^{2}_2$, and from the bottom sub-blocks, we have
$\gamma^{3}_1=\gamma^{1}_2$. From this, we can deduce that
$\gamma^{1}_1 =\gamma^{2}_2 = \gamma^{2}_1 = \gamma^{1}_2$ so that
$\gamma^{1}_1 =\gamma^{1}_2$ and similarly, $\gamma^{2}_1
=\gamma^{2}_2$ and $\gamma^{3}_1 =\gamma^{3}_2$ so that $\gamma_1=\gamma_2$.

So, if we set $A_1=M_1$ and $A_2=M_2$, we have that $G_1$ and $G_2$ are
isomorphic if and only if $\hat A_1$ is similar to $\hat A_2$.
This completes the proof that if the extended adjacency matrices
are seen to be equivalent as allocation cubes, then the graphs are isomorphic.
Therefore, we have shown a polynomial-time transformation from
\textsc{Graph Isomorphism} to \textsc{Cube Similarity (2-d)}.

Next, we show a polynomial-time transformation from
\textsc{Cube Similarity (2-d)} to \textsc{Graph Isomorphism}.  We reduce
\textsc{Cube Similarity (2-d)} to \textsc{Directed Graph Isomorphism}, which is
in turn reducible to
\textsc{Graph Isomorphism}~\cite{gare:gandj,hunt:hunt-rosenk77}.

Given two 0-1 matrices $M_1$ and $M_2$, we want to decide whether we
can find $(\gamma _1, \gamma _2)$ such that
$(\gamma _1, \gamma _2 ) (M_1) = M_2$.
We can assume that $M_1$ and $M_2$ are square matrices
and if not, pad with as many rows or columns filled with zeros
as needed. We want a reduction from this problem to
\textsc{Directed Graph Isomorphism}.
Consider the following matrices: \begin{singlespace} $\hat M_1 =  \begin{bmatrix}
 0 & M_1 \\
 0 & 0
\end{bmatrix} $\end{singlespace}
and \begin{singlespace} $\hat M_2 = \begin{bmatrix}
 0 & M_2 \\
 0 & 0
\end{bmatrix}$\end{singlespace}.
Both $\hat M_1$ and $\hat M_2$ can be considered as the adjacency
matrices of directed graphs $G_1$ and $G_2$.
Suppose that the graphs
are found to be isomorphic, then there is a permutation
$\gamma$ such that $(\gamma,\gamma)(\hat M_1) = \hat M_2$.
We can assume without loss of generality that $\gamma$ does not
permute rows or columns having only zeros across halves of
the adjacency matrices. On the other hand, rows containing
non-zero components cannot be permuted across halves.
Thus, we can decompose $\gamma$ into two disjoint permutations
$\gamma^1$ and $\gamma^2$ and hence
$(\gamma^1,\gamma^2)(M_1)=M_2,$
which implies $M_1\sim M_2$. On the other hand, if $M_1\sim M_2$,
then there is $(\gamma^1,\gamma^2)$ such that
$(\gamma^1,\gamma^2)(M_1) = M_2$ and we can choose
$\gamma$ as the direct sum of $\gamma^1$ and $\gamma^2$.
Therefore, we have found a reduction from
\textsc{Cube Similarity (2-d)} to \textsc{Directed Graph Isomorphism} and,
by transitivity, to \textsc{Graph Isomorphism}.

Thus, \textsc{Graph Isomorphism} and \textsc{Cube Similarity (2-d)} are mutually
reducible 
and hence \textsc{Cube Similarity (2-d)} is
\textsc{Graph Isomorphism}-complete.
 \end{proof}

\begin{rem}
If similarity between two $n\times n$ cubes can be decided in
time $c n^k$ for some positive integers $c$ and $k \geq 2$, then
graph isomorphism can be decided in O($n^k$) time.
\end{rem}

Since \textsc{Graph Isomorphism} has been reduced to a special
case of \textsc{Cube Similarity}, then
the general problem is at least as difficult as \textsc{Graph
Isomorphism}.  Yet we have seen no reason to believe the general
problem is harder (for instance, NP-complete).
We suspect that a stronger result may be possible; establishing
(or disproving) the following conjecture is left as an open problem.

\begin{conj}
The general \textsc{Cube Similarity} problem is
also \textsc{Graph Iso\-mor\-phism}-complete.
\end{conj}

\section{Computational Complexity of Optimal Normalization}
\label{section:generalcomplexity}

It appears that it is computationally intractable
to find a ``best'' normalization $\pi$
(i.e., $\pi$ minimizes cost per allocated cell $E(\pi(C))$)
given a cube $C$ and given the blocks' dimensions.
Yet, when suitable restrictions are imposed, a best normalization
can be computed (or approximated) in polynomial time.  This section
focuses on the effect of block size on intractability.

\subsection{Tractable Special Cases}

Our problem can be solved in polynomial time, if severe
restrictions are placed on the number of dimensions or
on block size.  For instance,
it is trivial to find a best normalization in 1-d.
Another trivial case arises when blocks are of size 1, since then normalization
does not affect storage cost.   Thus, any normalization is a
``best normalization.''  The situation is more interesting for
blocks of size 2; i.e., which have $m_i = 2$ for some $1 \leq i \leq d$
and $m_j = 1$ for $1 \leq j \leq d$ with $i \not = j$.
A best normalization can be found in polynomial time, based on
weighted-matching~\cite{gabo:edmonds-impl} techniques described
next.

\subsubsection{Using Weighted Matching}
\label{matching-volume-two}

Given a weighted undirected graph, the
\emph{weighted matching problem}  asks for an edge subset of
maximum or minimum total weight, such that no two edges
share an endpoint.  If the graph is complete,
has an even number of vertices, and has only positive edge
weights, then the maximum matching
effectively pairs up vertices.

\begin{figure}
\begin{center}\resizebox{9cm}{!}{\includegraphics{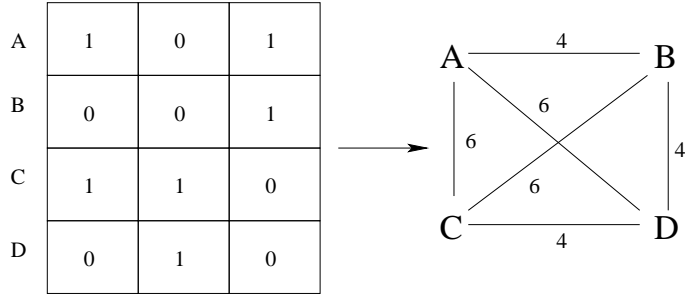}}\end{center}

\caption[Mapping a normalization problem to a
weighted matching problem on graphs.]{\label{matchfig} Mapping a normalization problem to a
weighted matching problem on graphs.  Rows are labeled and we try to
reorder them, given block dimensions $2 \times 1$ (where 2 is the
vertical dimension). In this example, optimal solutions include
$r_0,r_1,r_2,r_3$ and $r_2,r_3,r_1,r_0$. }
\end{figure}

For our problem, normalization's effect on dimension $k$,
for some $1 \leq k \leq d$, corresponds to rearranging the
order of the $n_k$ slices $C^{k}_v$, where $1 \leq v \leq n_k$.
In our case, we are using a block size of 2 for dimension $k$.
Therefore, once we have chosen two slices $C^{k}_v$ and $C^{k}_{v'}$
to be the first pair of slices, we will have formed the first
layer of blocks and have stored all allocated cells belonging to
these two slices.
The total storage cost of the cube is thus a sum, over all pairs
of slices, of the pairing-cost of the two slices composing the pair.
The order in which pairs are chosen is irrelevant: only the
actual matching of slices into pairs matters.
Consider Boolean vectors  $\mathbf{b} = \widehat{C}^{k}_v$ and
$\mathbf{b'} = \widehat{C}^{k}_{v'}$.  If
both $\mathbf{b}_i$ and $\mathbf{b'}_i$ are true,
then the $i^{th}$ block
in the pair is completely full and costs 2 to store. Similarly, if
exactly one of $\mathbf{b}_i$ and $\mathbf{b'}_i$ is true,
then the block is half-full.
Under our model, a half-full block also costs 2, but an empty block costs 0.
Thus, given any two slices, we can compute the cost of
pairing them by summing the storage costs of all these blocks.
If we identify each slice with a vertex of a complete
weighted graph, it is easy to form an instance of weighted
matching. (See Figure~\ref{matchfig} for an example.)
Fortunately, cubic-time algorithms exist for weighted
matching~\cite{NetworkFlows1993}, 
and $n_k$ is often small enough that cubic running time
is not excessive. Unfortunately, calculating the 
$n_k(n_k-1)/2$
edge weights is expensive; each involves two large
Boolean vectors with
$\frac{1}{n_{k}}\prod_{i=1}^d n_i$ elements,
for a total edge-calculation time of
$\Theta \left ( n_k \prod_{i=1}^d n_i \right )$.
Fortunately, this can be improved for sparse cubes.

In the 2-d case, given any two rows, for example \begin{singlespace}$r_1 =
\begin{bmatrix}0 & 0 & 1 & 1\end{bmatrix}$\end{singlespace}
and \begin{singlespace}$r_2 = \begin{bmatrix}0
& 1 & 0 & 1\end{bmatrix}$\end{singlespace}, then we can compute the total allocation
cost of grouping the two together as $2(\#r_1 +\#r_2 -
\mathit{benefit} )$ where $\mathit{benefit}$ is the number of
positions (in this case 1) where both $r_1$ and $r_2$ have allocated
cells.  (This \textit{benefit} records that one of the two allocated
values could be stored ``for free,'' were slices $r_1$ and $r_2$
paired.)

According to this formula, the cost of putting $r_1$ and $r_2$
together is thus $2(2+2-1)=6$.  Using this formula, we can improve
edge-calculation time when the cube is sparse.  To do so, for each of
the $n_k$ slices $C^{k}_v$, represent each allocated value by a
$d$-tuple $(i_1, i_2, \ldots , i_{k-1}, i_{k+1}, \ldots, i_{d}, i_k)$
giving its coordinates within the slice and labeling it with the
number of the slice to which it belongs.  Then sort these $\#C$ tuples
lexicographically, in O($\#C \log \#C$) time.  For example, consider
the following cube, where the rows have been labeled from $r_0$ to $r_5$ (
$r_i$ corresponds to $C^1_i$):
\[
\begin{bmatrix}
 r_0 & 0 & 0 & 0 & 0 \\
 r_1 & 1 & 1 & 0 & 1 \\
 r_2 & 1 & 0 & 0 & 0 \\
 r_3 & 0 & 1 & 1 & 0 \\
 r_4 & 0 & 1 & 0 & 0 \\
 r_5 & 1 & 0 & 0 & 1
\end{bmatrix}.
\]
We represent the allocated cells as \{$(0,r_1)$, $(1,r_1)$, $(3,r_1)$,
$(0,r_2)$, $(1,r_3)$, $(2,r_3)$, $(1,r_4)$, $(0,r_5)$, and
$(3,r_5)$\}. We can then sort these to get $(0,r_1)$, $(0,r_2)$,
$(0,r_5)$, $(1,r_1)$, $(1,r_3)$, $(1,r_4)$, $(2,r_3)$, $(3,r_1)$,
$(3,r_5)$.  This groups together allocated cells with corresponding
locations but in different slices. For example, two groups are
($(0,r_1)$, $(0,r_2)$,
$(0,r_5)$) and ($(1,r_1)$, $(1,r_3)$, $(1,r_4)$).  
Initialize the \textit{benefit} value associated to each edge to zero, and
next process each group.
Let $g$ denote the number of tuples in the current group, and
in $O(g^2)$ time examine all $\chooz{g}{2}$
pairs of slices $(s_1,s_2)$ in the group, and increment (by 1) the
\textit{benefit} of the graph edge $(s_1,s_2)$.
In our example, we would process the group ($(0,r_1)$, $(0,r_2)$, $(0,r_5)$)
and increment the \textit{benefit}s of edges ($r_1,r_2$), 
($r_2,r_5$), and ($r_1,r_5$). For group
($(1,r_1)$, $(1,r_3)$, $(1,r_4)$), we would increase the \textit{benefit}s
of edges ($r_1,r_3$), ($r_1,r_4$), and ($r_3,r_4$).
Once all $\#C$ sorted tuples have been processed, the eventual
weight assigned to edge $(v,w)$ is
$2(\#\hat{C}^k_v + \#\hat{C}^k_w - \mathit{benefit}(v,w))$.
In our example, we have that edge $(r_1,r_2)$ has a benefit of 1,
and so a weight of $2(\#r_1+\#r_2-\mathit{benefit})=2(3+1-1)=6$.

A crude estimate of the running time to process the groups
would be that each group is O($n_k$) in size, and there are
O($\#C$) groups, for a time of O($\#C n^2_k$). It can be
shown that time is maximized when the $\#C$ values are distributed
into $\#C/n_k$ groups of size $n_k$, leading to a time
bound of $\Theta(\#C n_k)$ for group processing, and
an overall edge-calculation time of $\#C (n_k + \log \#C)$.

\begin{thm}
The best normalization for blocks of size  
$\overbrace{1\times\ldots\times1}^{i}
\times2\times
\overbrace{1\ldots\times1}^{k-1-i}$
can be computed in
$O(n_k \times (n_1 \times n_2 \times \ldots \times n_d) + n_k^3)$ time.
\end{thm}

The improved edge-weight calculation (for sparse cubes) leads to the
following.

\begin{cor}
The best normalization for blocks of size
$\overbrace{1\times\ldots\times1}^{i}
\times2\times
\overbrace{1\ldots\times1}^{k-1-i}$
can be computed in 
$O(\#C (n_k + \log \#C)  + n_k^3)$ time.

\end{cor}

For more general
block shapes, this algorithm is no longer optimal but nevertheless
provides a basis for sensible heuristics.

\subsection{An NP-hard Case}

In contrast to the $1\times2$-block situation, we next
show that it is NP-hard to find the best normalization for $1\times3$ blocks.
\chainsaw{
The associated decision problem asks whether any normalization can store a given cube
within a given storage bound, assuming $1\times3$ blocks.   
}
We return to the general cost model from 
Section~\ref{section:block-coded} but choose $\alpha = 1/4$, as this
results in an especially simple situation where
a block with three allocated cells ($D=3$) stores 
each of them at a cost of 1, whereas a block with fewer
than three allocated cells stores each allocated cell at a cost of $3/2$.

The proof involves a reduction from the NP-complete problem
Exact 3-Cover (X3C),
a problem which gives a set $S$ and a set $\mathcal{T}$ of three-element
subsets of $S$.  The question, for X3C, is whether there is a 
$\mathcal{T}' \subseteq \mathcal{T}$ such that each $s \in S$ occurs in exactly one 
member of $\mathcal{T}'$~\cite{gare:gandj}.

We sketch the reduction next.  Given an instance of X3C, form an
instance of our problem by making a $|\mathcal{T}| \times |S| $ cube. For 
$s \in S$ and $T \in \mathcal{T}$, the cube has an allocated cell corresponding to $(T,s)$
if and only if $s \in T$.  Thus, the cube has $3 | \mathcal{T} |$ cells that need
to be stored.  The storage cost cannot be lower than $\frac{9 | \mathcal{T} | - |S|}{2}$
and this bound can be met if and only if the answer to the instance of X3C is
``yes.'' 
\chainsaw{ Indeed,
a normalization for $1\times3$ blocks can be viewed as
simply grouping the values of an attribute into triples.  Suppose the
storage bound is achieved, then at least $|S|$ cells would have to be
stored in full blocks. Consider some full block and note there are
only 3 allocated cells in each row, so all 3 of them must be chosen
(because blocks are $1\times3$).  But the three allocated cells in
a row can be mapped to a $T \in \mathcal{T}$. Choose it for
$\mathcal{T'}$.  None of these 3 cells' columns intersect any other
full blocks, because that would imply some other row had exactly the
same allocation pattern and hence represents the same $T$, which it
cannot.  So we see that each $s \in S$ (column) must intersect exactly
one full block, showing that $\mathcal{T'}$ is the cover we seek.

Conversely, suppose $\mathcal{T'}$ is a cover for X3C.  Order the elements
in $\mathcal{T'}$ arbitrarily as $T_0, T_1, \ldots, T_{|S|/3}$ and
use any normalization that puts first (in arbitrary order) 
the three $s\in T_0$,
then next puts the three $s \in T_1$, and so forth.  
The three allocated cells for each $T_i$ will be
together in a (full) block, giving
us at least the required ``space savings'' of $\frac{3}{2} |\mathcal{T'}| = |S|$.
} 

\begin{thm}
It is NP-hard to find the best normalization when $1\times3$ blocks are used.
\end{thm}

We conjecture that it is NP-hard to find the best normalization
whenever the block size is fixed at any size larger than 2.
\chainsaw{
A related 2-d problem that is NP-hard was discussed by 
Kaser~\cite{kase:compress-article}.
Rather than specify the block dimensions, this problem allows the
solution to specify how to divide each dimension into two ranges,
thus making four blocks in total (of possibly different shape)
.}

\section{Slice-Sorting Normalization for Quasi-Independent Attributes}
\label{section:slicesort}

In practice, whether or not a given cell is allocated may depend
on the corresponding attribute values independently
of each other. For example, if a store is closed on Saturdays
almost all year, a slice corresponding to {}``weekday=Saturday''
will be sparse irrespective of the other attributes.
In such cases, it is sufficient to normalize the data cube
using only an attribute-wise approach. Moreover, as we shall
see, one can easily compute the degree of independence of
the attributes and thus decide whether or not potentially
more expensive algorithms need to be used.

We begin by examining one of the simplest classes of normalization
algorithms, and we will assume $n$-regular data cubes for $n\geq 3$. We
say that a sequence of values $x_{1},\ldots,x_{n}$ is sorted in
increasing (respectively, decreasing) order if $x_{i}\leq x_{i+1}$
(respectively, $x_{i}\geq x_{i+1}$) for $i \in \lbrace 1,\ldots,n-1
\rbrace$.

Recall that $\widehat{C}_{v}^{j}$ is the Boolean array indicating
whether a cell is allocated or not in slice $C_v^j$.

\begin{alg} \label{algo:SliceSort} (Slice-Sorting Normalization)
Given an $n$-regular data cube $C$, then slices have
$n^{d-1}$ cells. 
 Given a fixed function $g:\lbrace \mathit{true},\mathit{false}
\rbrace^{n^{d-1}}\rightarrow \R$, then for each attribute $j$, we
\glossary{name={$g$},description={a slice-ranking function}}
compute the sequence $f_{v}^{j}=g(\widehat{C}_{v}^{j})$ for all attribute values
$v=1,\ldots,n$.  Let $\gamma^{j}$ be a permutation such that
$\gamma^{j}(f^{j})$ is sorted either in increasing or decreasing
order, then a slice-sorting normalization is  
$( \gamma^{1},\ldots, \gamma^{d} )$.
\end{alg}

\newcommand{\simm}{\simeq}  

Algorithm~\ref{algo:SliceSort} has time complexity $O(dn^d+d n \log n)$. 
We can precompute the aggregated values $f_{v}^{j}$
and speed up normalization to $O(d n \log(n))$. 
It does not produce
a unique solution given a function $g$ because there could
be many different valid ways to sort.
A normalization
$\varpi= ( \gamma^{1},\ldots, \gamma^{d} )$ is a
\glossary{name={$\varpi$},description={a slice-sorting normalization}}
\emph{solution to the slice-sorting problem} if it provides a valid sort for
the slice-sorting problem stated by Algorithm~\ref{algo:SliceSort} .
Given a data cube $C$, denote the set of all solutions to the slice-sorting problem
by $\mathcal{S}_{C,g}$. 
\glossary{name={$\mathcal{S}_{C,g}$},description={slice-sorting solution set}}
Two functions $g_{1}$ and
$g_{2}$ are \emph{equivalent}
with respect to the slice-sorting problem if
$\mathcal{S}_{C,g_{1}}=\mathcal{S}_{C,g_{2}}$ for all cubes $C$ and
we write $g_{1}\simm g_{2}$ . We can characterize such equivalence classes using monotone
functions. Recall that a function $h:\R \rightarrow \R$
is strictly monotone nondecreasing (respectively, nonincreasing)
if $x < y$ implies $h(x) < h(y)$ (respectively, $h(x) > h(y)$).

\chainsaw{
An alternative definition is that $h$ is monotone if,
whenever $x_{1},\ldots,x_{n}$
is a sorted list, then so is $h(x_{1}),\ldots,h(x_{n})$.
This second definition can be used to prove the existence of
a monotone function as the next proposition shows.

\begin{prop} \label{proposition:existencemonotone} For a fixed integer $n\geq 3$ and two
functions $\omega _{1},\omega _{2}:D\rightarrow \R$ where $\mathcal{D}$ is a set
with an order relation, if for all sequences $x_{1},\ldots,x_{n} \in
\mathcal{D}$, $\omega _{1}(x_{1}),\ldots,\omega _{1}(x_{n})$ is sorted if and
only if $\omega _{2}(x_{1}),\ldots,\omega _{2}(x_{n})$ is sorted, then
there is a monotone function $h:\R \rightarrow \R$ such that
$\omega _{1}=h\circ \omega _{2}$.
\end{prop}
\begin{proof}  The proof is constructive. Define $h$ over the image
of $\omega _{2}$ by the formula
$$h(\omega _{2}(x))=\omega _{1}(x).$$
To prove that $h$ is well defined, we have to show that whenever
$\omega _{2}(x_{1}) = \omega _{2}(x_{2})$ then
$\omega _{1}(x_{1}) = \omega _{1}(x_{2})$. Suppose that this
is not the case, and without loss of generality, let
$\omega _{1}(x_{1}) < \omega _{1}(x_{2})$. Then there is
$x_{3} \in \mathcal{D}$ such that
$\omega _{1}(x_{1})\leq \omega _{1}(x_{3})\leq   \omega _{1}(x_{2})$
or
$\omega _{1}(x_{3})\leq \omega _{1}(x_{1})$
or
$\omega _{1}(x_{2})\leq \omega _{1}(x_{3})$.
In all three cases, because of the equality between
$\omega _{2}(x_{1})$ and $\omega _{2}(x_{2})$,
any ordering of $\omega _{2}(x_{1}), \omega _{2}(x_{2}), \omega _{2}(x_{3})$
is sorted whereas there is always one non-sorted sequence using $ \omega _{1}$.
There is a contradiction, proving that $h$ is well defined.

For any sequence $x_{1},x_{2},x_{3}$ such that $\omega _{2}(x_{1}) <
\omega _{2}(x_{2}) < \omega _{2}(x_{3})$, then we must either have
$\omega _{1}(x_{1}) \leq \omega _{1}(x_{2}) \leq \omega _{1}(x_{3})$
or $\omega _{1}(x_{1}) \geq \omega _{1}(x_{2}) \geq \omega
_{1}(x_{3})$ by the conditions of the proposition. In other words, for
$x<y<z$, we either have $h(x)\leq h(y)\leq h(z)$ or $h(x)\geq h(y)\geq
h(z)$ thus showing that $h$ must be monotone.
 \end{proof}

\begin{prop} \label{proposition:equivalenceofslicesort}
Given two functions $g_{1},g_{2}:\lbrace \mathit{true},\mathit{false} \rbrace^{S}\rightarrow \R$,
we have that
$$\mathcal{S}_{C,g_{1}}=\mathcal{S}_{C,g_{2}}$$ for all data cubes $C$
if and only if there exist a monotone function $h:\R \rightarrow \R$
such that $g_{1}=h\circ g_{2}$.
\end{prop}
\begin{proof}  Assume 
there is $h$ such that $g_{1}=h\circ g_{2}$,
and consider $\varpi = ( \gamma^{1},\ldots,\gamma^{d} ) \in \mathcal{S}_{C,g_{1}}$ for any data
cube $C$, then $\gamma^{j}(g_{1}(\widehat{C}_{v}^{j}))$ is sorted over index $v\in \lbrace 1,\ldots,n \rbrace$
for all attributes $j=1,\ldots,n$ by definition of $\mathcal{S}_{C,g_{1}}$.
Then $\gamma^{j}(h(g_{1}(\widehat{C}_{v}^{j})))$ must also be sorted
over $v$ for all $j$, since  monotone functions
preserve sorting. Thus $\varpi \in \mathcal{S}_{C,g_{2}}$.

One the other hand, if $\mathcal{S}_{C,g_{1}}=\mathcal{S}_{C,g_{2}}$ for all data cubes $C$,
then $h$ exists by  Proposition~\ref{proposition:existencemonotone}.
  \end{proof}

} 

\verbose{Consider a general normalization algorithm, $\zeta$, which is not necessarily 
slice sorting, and which, given a cube $C$,
outputs exactly one normalization, $\zeta_C$. The newly normalized
cube is $\zeta_C (C)$. Of course, we can normalize
the normalized cube. However, if $\zeta_C (C) = \zeta_{\zeta_C(C)} (\zeta_C(C))$
--- equivalently, $\zeta_{\zeta_C(C)} = I$ ---
for all cubes $C$,
then $\zeta$ terminates after one normalization and we say the
normalization algorithm is \emph{strictly stable}.
 Stability tells us
that there is no need to apply the algorithm more than once,
as we would keep on getting the same result.
This is not the only desirable property, however.
What if the original cube had been normalized differently?
Two users may feed $\zeta$ the same cube normalized
differently, and we would want the same output. Thus, 
we want $\zeta_C (C) = \zeta_{\varpi(C)} (\varpi(C))$ or,
in other words, 
$\zeta_C = \zeta_{\varpi(C)} \circ \varpi$. If this is
true for all cubes $C$, we say the normalization algorithm
is \emph{strictly strongly stable}. Strict
strong stability implies
strict stability as can be seen by choosing $\varpi=\zeta_C$.
The converse is not true. Consider a slice-sorting
algorithm: the order of the sort will depend
on the order of the input data. Yet, some sorting algorithms
are stable: if the data is already sorted, they will act
as the identity transformation. This stability can easily
be achieved by adding a check before the sort to verify if
the data is already sorted.
 In practice, it might be
difficult to achieve strict strong stability, and
so, in the context of slice sorting algorithms, we consider
somewhat weaker concepts.}

A slice-sorting algorithm is \emph{stable} if the
normalization of a normalized cube can be chosen to be the identity,
that is if $\varpi\in \mathcal{S}_{C,g}$ then $I\in
\mathcal{S}_{\varpi(C),g}$ for all $C$. The algorithm is
\emph{strongly stable} if for any normalization $\varpi$,
$\mathcal{S}_{\varpi(C),g} \circ \varpi =\mathcal{S}_{C,g}$ for all
$C$. 
Strong stability means that the resulting normalization
does not depend on the initial normalization.
This is a desirable property because
data cubes are often normalized arbitrarily at construction
time.  Notice that strong stability implies stability: choose $\varpi
\in \mathcal{S}_{C,g}$. Then there must exist $\zeta \in
\mathcal{S}_{\varpi(C),g} $ such that $\zeta \circ \varpi = \varpi$
which implies that $\zeta$ is the identity.

\begin{prop}Stability implies strong stability for 
slice-sorting
algorithms and so, strong stability $\Leftrightarrow$ stability.
\end{prop}
\begin{proof}
Consider a slice-sorting algorithm, based on $g$, that is stable.  Then
by definition
\begin{equation}
\varpi \in \mathcal{S}_{C,g} \Rightarrow I \in \mathcal{S}_{\varpi(C),g} \label{eqss1}
\end{equation}
 for all $C$.
Observe that the converse is true as well, that is,
\begin{equation}
I \in \mathcal{S}_{\varpi(C),g}  \Rightarrow \varpi \in \mathcal{S}_{C,g}. \label{eqss2}
\end{equation}

Hence we have that $\varpi_1 \circ \varpi \in \mathcal{S}_{C,g}$ implies that
$I \in \mathcal{S}_{\varpi_1(\varpi(C)),g}$ by Equation~\ref{eqss1} and so,
by Equation~\ref{eqss2}, $\varpi_1 \in \mathcal{S}_{\varpi(C),g}$. Note
that given any $\varpi$, all elements of $\mathcal{S}_{C,g}$
can be written as $\varpi_1 \circ \varpi$ because permutations are invertible.
Hence,
given $\varpi_1 \circ \varpi \in  \mathcal{S}_{C,g}$ we have
$\varpi_1 \in  \mathcal{S}_{\varpi(C),g}$ and so
$\mathcal{S}_{C,g} \subset \mathcal{S}_{\varpi(C),g} \circ \varpi$.

 On the other hand, given
$\varpi_1 \circ \varpi \in \mathcal{S}_{\varpi(C),g}  \circ \varpi$,
we have that $\varpi_1  \in \mathcal{S}_{\varpi(C),g}$ by cancellation,
hence $I \in \mathcal{S}_{\varpi_1 (\varpi(C)),g}$ by Equation~\ref{eqss1},
and then $\varpi_1 \circ \varpi \in \mathcal{S}_{C,g}$ by Equation~\ref{eqss2}.
Therefore, $\mathcal{S}_{\varpi(C),g}  \circ \varpi \subset \mathcal{S}_{C,g}$.
\end{proof}

Define $\tau :\lbrace \mathit{true},\mathit{false} \rbrace^{S}\rightarrow \R$
as the number of $\mathit{true}$ values in the argument. In effect, $\tau$ 
\glossary{name={$\tau$},description={counts the number of allocated cells}}
counts the number of allocated cells: $\tau(\widehat{C}_{v}^{j}) = \#\widehat{C}_{v}^{j}$ for any slice $\widehat{C}_{v}^{j}$.
If the slice $\widehat{C}_{v}^{j}$ is normalized,
$\tau$ remains constant: 
$\tau(\widehat{C}_{v}^{j})=\tau \left (\varpi \left ( \widehat{C}_{v}^{j} \right  ) \right)$
for all normalizations $\varpi$. Therefore $\tau$ leads to a strongly stable slice-sorting
algorithm.
 The converse is also true if $d=2$%
\chainsaw{
, that is, if the slice is one-dimensional, then 
if $$h(\widehat{C}_{v}^{j})=h \left (\varpi \left ( \widehat{C}_{v}^{j} \right  ) \right)$$
for all normalizations $\varpi$ then $h$ can only depend on the number
of allocated ($\mathit{true}$) values in the slice since it fully characterizes
the slice up to normalization. For  the general case ($d>2$), the converse is
not true since the number of allocated values is not enough to
characterize the slices up to normalization. For example, one could
count how many sub-slices along a chosen second attribute have
no allocated value.
} 

\chainsaw{ A function $g$ is \emph{symmetric} if $g\circ
\varpi \simm g$ for all normalizations $\varpi$. The following
proposition shows that up to a monotone function, strongly stable
slice-sorting algorithms are characterized by \emph{symmetric
functions}.

\begin{prop} 
\label{proposition:characterizationofstrongstab}
A slice-sorting algorithm based on a function $g$
is strongly stable if and only if for any normalization $\varpi$,
there is a monotone function $h:\R \rightarrow \R$
such that 
\begin{equation}
g \left (\varpi \left ( \widehat{C}_{v}^{j} \right  ) \right)
= h \left( g(\widehat{C}_{v}^{j}) \right) \label{eq:htovarpi}
\end{equation}
for all attribute values $v=1,\ldots,n$ of all attributes $j=1,\ldots,d$.
In other words, it is strongly stable if and only if $g$ is symmetric.
\end{prop}
\begin{proof}  By Proposition~\ref{proposition:equivalenceofslicesort},
Equation~\ref{eq:htovarpi} is sufficient for strong stability. On the
other hand, suppose that the slice-sorting algorithm is strongly stable
and that there does not exist a strictly monotone function $h$ satisfying
Equation~\ref{eq:htovarpi}, then by 
Proposition~\ref{proposition:existencemonotone}, there must be
a sorted sequence $g(\widehat{C}_{v_{1}}^{j}),g(\widehat{C}_{v_{2}}^{j}),
g(\widehat{C}_{v_{3}}^{j})$
such that 
$g \left (\varpi \left ( \widehat{C}_{v_{1}}^{j} \right  ) \right),
g \left (\varpi \left ( \widehat{C}_{v_{2}}^{j} \right  ) \right),
g \left (\varpi \left ( \widehat{C}_{v_{3}}^{j} \right  ) \right)$ is not
sorted. Because this last statement contradicts strong stability, we have
that Equation~\ref{eq:htovarpi} is necessary.
\end{proof} 
}  

\begin{lem} A slice-sorting algorithm based on a function $g$
is strongly stable if  $g=h \circ \tau$ for some function $h$.
For 2-d cubes, the condition
is necessary.
\end{lem}

In the above lemma, whenever $h$ is strictly monotone, then $g \simm \tau$
and we call this class of slice-sorting algorithms \emph{Frequency
Sort}~\cite{kase:compress-tr}. We will show that we can estimate
\textit{a priori} the efficiency of this class 
(see Theorem~\ref{theorem:bigbadtheorem}).
\verbose{For Frequency Sort,
 we can precompute the number of
 allocated cells per slice $\#\widehat{C}_{v}^{j}$ as the data
 cube is constructed, thus speeding up the normalization
 phase and making it $O(d n \log(n))$.
}

It is useful to consider a data cube as a probability distribution
in the following sense: given a data cube $C$, let the
\emph{joint probability distribution} $\Psi$ over the same
\glossary{name={$\Psi$},description={the joint probability distribution: $\Psi_{i_{1},\ldots,i_{n}}$ is the probability that cell ($i_{1},\ldots,i_{n}$) is chosen if we are to choose one allocated cell in the cube}}
$n^d$ set of indices be
$$\Psi_{i_{1},\ldots,i_{n}} = \left \lbrace \begin{array}{cc}
 1/\#C & \mbox{if~}C_{i_{1},\ldots,i_{n}}\neq 0  \\ 
 0 & \mbox{otherwise}
\end{array} \right. .$$
The underlying probabilistic model is that 
allocated cells are uniformly likely to be picked whereas
unallocated cells are never picked.
Given an attribute $j\in \lbrace 1,\ldots, d \rbrace$,
consider the number of allocated slices in slice $C_v^j$,  $\#\widehat{C}_{v}^{j}$, for $v\in \lbrace 1,\ldots, n \rbrace$:
we can define a \emph{probability distribution} $\varphi^{j}$ along attribute $j$
\glossary{name={$\varphi^{j}$},description={probability of choosing a cell in one of the slices of dimension $j$ if we are to choose one allocated cell in the cube, defined as $\varphi_{v}^{j}=\frac{\#\widehat{C}_{v}^{j}}{\#C}$}}
as $\varphi_{v}^{j}=\frac{\#\widehat{C}_{v}^{j}}{\#C}$.
From these $\varphi^{j}$ for all $j\in \lbrace 1,\ldots, d \rbrace$,
we can define the \emph{joint independent probability distribution} $\Phi$ as
\glossary{name={$\Phi$},description={joint independent distribution: $\Phi_{i_{1},\ldots,i_{d}} = \prod_{j=1}^{d} \varphi_{i_{j}}^{j}$}}
$\Phi_{i_{1},\ldots,i_{d}} = \prod_{j=1}^{d} \varphi_{i_{j}}^{j},$
or in other words $\Phi=\varphi^{0}\otimes \ldots \otimes\varphi^{d-1}$.
Examples are given in Table~\ref{cap:probadistributions}.

\begin{table}
\caption{\label{cap:probadistributions}
Examples of 2-d data cubes and their probability distributions.
}
{
\renewcommand{\baselinestretch}{1.2} \footnotesize   
\begin{center}
\begin{tabular}{|c|c|c|}
\hline
Data Cube & Joint Prob. Dist. & Joint Independent 
 Prob. Dist. \\
\hline
$ \begin{matrix}
1 & 0 & 1 & 0 \\
0 & 1 & 0 & 1 \\
1 & 0 & 1 & 0 \\
0 & 1 & 0 & 1
\end{matrix}$
&
$\begin{matrix}
\frac{1}{8} & 0           & \frac{1}{8} & 0           \\
0           & \frac{1}{8} & 0           & \frac{1}{8} \\
\frac{1}{8} & 0           & \frac{1}{8} & 0           \\
0           & \frac{1}{8} & 0           & \frac{1}{8}
\end{matrix}$
 &
$\begin{matrix}
\frac{1}{16} & \frac{1}{16} & \frac{1}{16} & \frac{1}{16} \\
\frac{1}{16} & \frac{1}{16} & \frac{1}{16} & \frac{1}{16} \\
\frac{1}{16} & \frac{1}{16} & \frac{1}{16} & \frac{1}{16} \\
\frac{1}{16} & \frac{1}{16} & \frac{1}{16} & \frac{1}{16}
\end{matrix}$
  \\

\hline
$ \begin{matrix}
1 & 0 & 0 & 0 \\
0 & 1 & 0 & 0 \\
0 & 1 & 1 & 0 \\
0 & 0 & 0 & 0
\end{matrix}$
&
 $\begin{matrix}
\frac{1}{4} & 0           & 0           & 0 \\
0           & \frac{1}{4} & 0           & 0 \\
0           & \frac{1}{4} & \frac{1}{4} & 0 \\
0           & 0           & 0           & 0
\end{matrix}$
 &
 $\begin{matrix}
\frac{1}{16} & \frac{1}{8} & \frac{1}{16} & 0 \\
\frac{1}{16} & \frac{1}{8} & \frac{1}{16} & 0 \\
\frac{1}{8}  & \frac{1}{4} & \frac{1}{8}  & 0 \\
0            & 0           & 0            & 0
\end{matrix}$
  \\
\hline
\end{tabular}
\end{center}
}
\end{table}

Given a joint probability distribution $\Psi$ and the number of allocated
cells $\#C$, we can build an \emph{allocation cube} $A$ by
\glossary{name={allocation cube},description={cube storing values between 0 and 1 indicating how likely it is that the cell is allocated}}
computing $ \Psi \times \#C$. Unlike a data cube, an allocation cube stores values
between 0 and 1 indicating how likely it is that the cell be allocated. 
 If we start from a data cube $C$
and compute its joint probability distribution and from it, its
allocation cube, we get a cube containing only 0's and 1's depending
on whether or not the given cell is allocated (1 if allocated, 0
otherwise) and we say we have the \emph{strict allocation cube} of the
\glossary{name={strict allocation cube},description={cube storing values 0 and 1 
indicating whether a cell is allocated or not}}
data cube $C$.  For an allocation cube $A$, we define $\#A$ as the sum
of all cells.  We define the normalization of an allocation cube in
the obvious way.  The more interesting case arises when we consider
the joint independent probability distribution: its allocation cube
contains 0's and 1's but also intermediate
values. Given an arbitrary allocation cube $A$ and another allocation
cube $B$, $A$ is \emph{compatible} with $B$ if any non-zero cell in $B$ has 
a value greater than the corresponding cell in $A$ and if all non-zero cells in $B$ are non-zero in $A$.
\glossary{name={compatible},description={$A$ is \emph{compatible} with $B$ if any 
non-zero cell in $B$ has a value greater than the corresponding cell in $A$  
and if all non-zero cells in $B$ are non-zero in $A$}}
We say that $A$ is \emph{strongly compatible} with $B$ if, in
addition to being compatible with $B$,  all
non-zero cells in $A$ are non-zero in $B$ 
\glossary{name={strongly compatible},description={$A$ is \emph{strongly compatible} with $B$ if $A$ is compatible with $B$ and
$A$'s non-zero cells are non-zero in $B$ }}
Given an allocation cube $A$ compatible with $B$, we can define the strongly
\glossary{name={$A_B$},description={given $A$ compatible with $B$, $A_B$ is a 
strongly compatible copy of $A$ where all zero cells in $B$ are set to zero}}
compatible allocation cube $A_{B}$ as
$${A_{B}}_{i_{1},\ldots,i_{d}}=\left \lbrace
\begin{array}{cc}
A_{i_{1},\ldots,i_{d}} &  \mbox{~if~} B_{i_{1},\ldots,i_{d}}\neq 0 \\ 
0 & \mbox{otherwise}   
\end{array} \right. $$
and we denote the remainder by $A_{B^{c}}= A - A_{B}$. 
\glossary{name={$A_{B^{c}}$},description={$A_{B^{c}}= A - A_{B}$}}
The following
result is immediate from the definitions.

\begin{lem}  \label{lemma:trivialainlamda} Given a data cube $C$ and its  
joint independent probability distribution $\Phi$, let $A$ be the 
allocation cube of $\Phi$, then we have
$A$ is compatible with $C$. Unless $A$ is also the strict 
allocation cube of $C$, $A$ is not strongly compatible with $C$.
\end{lem}

We can compute $H(A)$, the HOLAP cost of an allocation cube $A$, by looking at
each block. The cost of storing a block densely is still  $M
=m_{1}\times\ldots\times m_{d}$ whereas the cost of storing it
sparsely is $(d/2+1)\hat D$ where $\hat D$ is the sum of the  
0-to-1 values stored in the corresponding block. As before, a block is
stored densely when $\hat D \geq \frac{M}{(d/2+1)}$. When $B$ is
the strict allocation cube of a cube $C$, then $H(C)=H(B)$
immediately. If $\# A = \# B$ and $A$ is compatible with $B$, then $H(A)\geq
H(B)$ since the number of dense blocks can only be less. Similarly,
since $A$ is strongly compatible with $B$, $A$ has the set of
allocated cells as $B$ but with lesser values.  Hence $H(A)\leq H(B)$.

\begin{lem} \label{lemma:allocationcube} Given a data cube $C$ and its strict
allocation cube $B$, for all allocation cubes $A$ compatible with $B$ such
that $\# A = \# B$, we have $H(A)\geq H(B)$. On the other hand, if
$A$ is strongly compatible with $B$ but not necessarily $\# A = \# B$, then
$H(A)\leq H(B)$.
\end{lem}

A corollary of Lemma~\ref{lemma:allocationcube} is that the
joint independent probability distribution gives a bound on the HOLAP
cost of a data cube.
\begin{cor} \label{corollary:normalizationindependence} The allocation cube $A$ of
the joint independent probability distribution $\Phi$ of a data cube $C$
satisfies $H(A) \geq H(C)$.\end{cor}

Given a data cube $C$, consider a normalization $\varpi$ such that
$H(\varpi (C))$ is minimal and $\mathit{fs} \in \mathcal{S}_{C,\tau}$.
\glossary{name={$\mathit{fs}$},description={a frequency sort normalization}}
Since $H( \mathit{fs}(C)) \leq H( \mathit{fs}(A))$ by
Corollary~\ref{corollary:normalizationindependence} and
$H(\varpi(C)) \geq \#C$ by our cost model, then
\[H( \mathit{fs}(C)) - H(\varpi(C)) \leq H( \mathit{fs}(A)) - \#C .\]
In turn, $H( \mathit{fs}(A))$ may be estimated using only
the attribute-wise frequency distributions and thus we may have a
fast estimate of $H( \mathit{fs}(C)) - H(\varpi(C))$.
Also, because joint independent probability distributions
are separable, Frequency Sort is optimal over them.

\begin{prop}\label{proposition:ind}
Consider a data cube $C$ and the allocation cube $A$ of its joint independent probability
distribution. A Frequency Sort normalization $\mathit{fs}
\in \mathcal{S}_{C,\tau}$ is optimal over joint independent probability distributions
( $H(\mathit{fs}(A))$ is \textbf{minimal} ).
\end{prop}
\begin{proof}
In what follows, we consider
only allocation cubes from independent probability distributions
and proceed by induction.
Let $\hat D$ be the sum of cells
in a block and let $F_A(x)=\#(\hat D>x)$ and $f_A(x)=\#(\hat D=x)$
denote, respectively, the number of blocks where the count is greater  than (or equal to) $x$ for allocation cube $A$.

Frequency Sort is clearly optimal over any one-dimensional cube $A$
in the sense that in minimizes the HOLAP cost.
In fact, Frequency Sort maximizes $F_A(x)$, which is a stronger condition ($F_{fs(A)} (x) \geq F_A (x)$).

Consider two allocation cubes $A_1$ and $A_2$ and their product
$A_1 \otimes A_2$. Suppose that Frequency Sort is an optimal normalization
for both $A_1$ and $A_2$. Then the following argument shows that it must be so for $A_1 \otimes A_2$.
Block-wise, the sum of the cells in $A_1 \otimes A_2$, is given
by $\hat D=\hat D_1 \times \hat D_2$ where $\hat D_1$
and $\hat D_2$ are respectively the sum of cells in
$A_1$ and $A_2$ for the corresponding blocks.

We have that
\[F_{A_1 \otimes A_2} (x) = \sum_y f_{A_1}(y) F_{A_2}(x/y) = \sum_y F_{A_1} (x/y) f_{A_2}(y) \]
and $\mathit{\mathit{fs}}(A_1 \otimes A_2)=\mathit{fs}(A_1) \otimes \mathit{fs}(A_2)$.
By the induction hypothesis, $F_{\mathit{fs}(A_1)}(x)\geq F_{A_1}(x)$
and so $\sum_y F_{A_1} (x/y) f_{A_2}(y)\leq \sum_y F_{\mathit{fs}(A_1)} (x/y) f_{A_2}(y)$. But
we can also repeat the argument by symmetry
\[\sum_y F(\mathit{fs}(A_1)) (x/y) f_{A_2}(y)= \sum_y f_{\mathit{fs}(A_1)} (y) F_{A_2}(x/y)\leq \sum_y f_{\mathit{fs}(A_1)} (y) F_{\mathit{fs}(A_2)}(x/y)\]
and so $F_{A_1 \otimes A_2} (x) \leq F_{\mathit{fs}(A_1 \otimes A_2)} (x)$. The result then follows by induction.
\end{proof}

There is an
even simpler way to estimate $H( \mathit{fs}(C)) - H(\varpi(C))$ and thus
decide
whether Frequency Sorting is sufficient as
Theorem~\ref{theorem:bigbadtheorem} shows %
\chainsaw{(see Table~\ref{cap:bigbadtheoremexamples} for examples)}. It
should be noted that we give an estimate valid independently of the
dimensions of the blocks; thus, it is necessarily suboptimal.

\begin{thm}\label{theorem:bigbadtheorem} Given a data cube $C$, let $\varpi$ be an optimal normalization
and $\textit{fs}$ be a Frequency Sort normalization, then
$$H( fs(C)) - H(\varpi(C)) \leq \left (\frac{d}{2} + 1 \right)  ( 1 -   \Phi  \cdot B ) \#C $$
where $B$ is the strict allocation cube of $C$ and $\Phi$ is the
joint independent probability distribution. The symbol $\cdot$ denotes the scalar product defined in
the usual way.
\end{thm}
\chainsaw{
\begin{proof}  Let $A$ be the allocation cube of the joint independent
probability distribution. We
use the fact that
 $$H( \textit{fs}(C)) - H(\varpi(C)) \leq H( \mathit{fs}(A)) - H(\varpi(C)).$$
We have that $\mathit{fs}$ is an optimal normalization over
joint independent probability distribution by Proposition~\ref{proposition:ind}
so that $H( \mathit{fs}(A)) \leq H( \varpi(A))$.
Also $H(\varpi(C))=H(\varpi(B))$ by  definition so that
\begin{eqnarray*}
H( \mathit{fs}(C)) - H(\varpi(C)) &\leq & H( \varpi(A)) - H(\varpi(B)) \\
                                     &\leq & H( \varpi(A_{B})) + H( \varpi(A_{B^{c}})) - H(\varpi(B)) \\
                                     &\leq &  H( \varpi(A_{B^{c}}))
\end{eqnarray*}
since $ H( \varpi(A_{B}))  - H(\varpi(B)) \leq 0$ by Lemma~\ref{lemma:allocationcube}.

Finally, we have that \[H( \varpi(A_{B^{c}})) \leq \left  (\frac{d}{2} + 1 \right) \#A_{B^{c}}\]
and $ \#A_{B^{c}} =  ( 1 -   \Phi  \cdot B ) \# C$.
 \end{proof}
}

This theorem says that $ \Phi \cdot B$ gives a rough measure of how
well we can expect Frequency Sort to perform over all block
dimensions: when $\Phi \cdot B$ is very close to 1, we need not use
anything but Frequency Sort whereas when it gets close to 0, we can
expect Frequency Sort to be less efficient.  We call this coefficient
the \emph{Independence Sum}.

Hence, if the ROLAP storage cost is denoted by
$\mathit{rolap}$, the optimally normalized
block-coded cost by $\mathit{optimal}$,
and the Independence Sum
by $\mathit{IS}$,  \glossary{name={$\mathit{IS}$},description={Independence Sum: a measure of the statistical independence between the dimensions of a cube}}
we have the relationship
\[\mathit{rolap} \geq \mathit{optimal} + (1-\mathit{IS})
\mathit{rolap} \geq \mathit{fs} \geq \mathit{optimal}\]
where $\mathit{fs}$ is the block-coded cost using
Frequency Sort as a normalization algorithm.

\begin{table}\caption[Given data cubes, we give lowest possible HOLAP cost $H(\varpi(C))$.]{\label{cap:bigbadtheoremexamples}
Given data cubes, we give lowest possible HOLAP cost $H(\varpi(C))$
using $2\times 2$ blocks, and an example of a Frequency Sort HOLAP cost $H(\mathit{fs}(C))$
plus the independence product  $\Phi  \cdot B$ and the bound from
theorem~\ref{theorem:bigbadtheorem} for the lack of optimality of
Frequency Sort.
}
\begin{center}
\begin{singlespace}\footnotesize
\begin{tabular}{|c|c|c|c|c|}
\hline data cube $C$ & $H(\varpi(C))$ &  $H(\mathit{fs}(C))$ & $\Phi  \cdot B$ & $\left(\frac{d}{2} + 1 \right)  ( 1 -   \Phi  \cdot B )  \#C $ \\
\hline
$ \begin{matrix}
1 & 0 & 1 & 0 \\
0 & 1 & 0 & 1 \\
1 & 0 & 1 & 0 \\
0 & 1 & 0 & 1
\end{matrix}$
& 8 & 16 & $\frac{1}{2}$ & 8 \\
\hline
$ \begin{matrix}
1 & 0 & 0 & 0 \\
0 & 1 & 0 & 0 \\
0 & 1 & 1 & 0 \\
0 & 0 & 0 & 0
\end{matrix}$
& 6 & 6  & $\frac{9}{16}$ & $\frac{7}{2}$ \\
\hline
$ \begin{matrix}
 1 & 0 & 1 & 0 \\
 0 & 1 & 1 & 1 \\
 1 & 1 & 1 &  0 \\
 0 & 1 & 0 & 1
\end{matrix}$
& 12  & 16  & $\frac{17}{25}$ & $\frac{32}{5}$ \\
\hline
$ \begin{matrix}
 1 & 0 & 0 & 0 \\
 0 & 1 & 0 & 0 \\
 0 & 0 & 1 & 0 \\
 0 & 0 & 0 & 1
\end{matrix}$
& 8 & 8 & $\frac{1}{4}$ & $6$  \\
\hline
\end{tabular}
\end{singlespace}
\end{center}
\end{table}

\section{Heuristics}
\label{section:heuristics}

Since many practical cases appear intractable, we must resort to
heuristics when the Independence Sum is small.  We have experimented
with several different heuristics, and we can categorize possible
heuristics as block-oblivious versus block-aware, dimension-at-a-time
or holistic, orthogonal or not.

\emph{Block-aware} heuristics use information about the shape and
positioning of blocks. In contrast,
Frequency Sort (FS) is an example of a \emph{ block-oblivious}
heuristic: it makes no use of block information (see Fig.~\ref{algo:FS}).  
\chainsaw{
Overall, block-aware heuristics
should be able to obtain better performance when the block size is
known, but may obtain poor performance when the block size 
used does not match the block size assumed during normalization.
The block-oblivious heuristics should be more robust.
}

\begin{figure}
{\singlespace
 \begin{algorithmic}
 \STATE \textbf{input} a cube $C$
 \FORALL {dimensions $i$}
 	\FORALL {attribute values $v$}
 		\STATE Count the number of allocated cells in corresponding slice (value of $\# \widehat{C}^i_v$)
	\ENDFOR
  	\STATE sort the attribute values $v$ according to $\# \widehat{C}^i_v$
   \ENDFOR
 \end{algorithmic}}
\caption{\label{algo:FS}Frequency Sort (FS) Normalization Algorithm}

 \end{figure}

All our heuristics reorder one dimension at a time, as opposed to a
``holistic'' approach when several dimensions are simultaneously
reordered.  In some heuristics, the permutation chosen for one
dimension does not affect which permutation is chosen for another
dimension.  Such heuristics are \emph{orthogonal}, and all the
strongly stable slice-sorting algorithms in
Section~\ref{section:slicesort} are examples. Orthogonal heuristics
can safely process dimensions one at a time, and in any order.
With non-orthogonal heuristics that process one dimension
at a time, we typically process all dimensions once, 
and repeat until some stopping condition  is met.

\subsection{Iterated Matching heuristic}

We have already shown that the weighted-matching algorithm can produce
an optimal normalization for blocks of size 2 (see
Section~\ref{matching-volume-two}). The Iterated Matching (IM)
heuristic processes each dimension independently, behaving each time as if
the blocks consisted of two cells aligned with the current dimension (see
Fig.~\ref{algo:IM}).  \glossary{name={IM},description={Iterated
Matching heuristic}} Since it tries to match slices two-by-two so as to
align many allocated cells in blocks of size 2, it should
perform well over 2-regular blocks. It processes each dimension
exactly once because it is orthogonal.

This algorithm is better explained using an example.
Applying this algorithm along the rows of the cube in Fig.~\ref{matchfig} (see page~\pageref{matchfig})
amounts to building the graph in the same figure and solving the weighted-matching problem over this graph.
The cube would then be normalized to
 \[\begin{bmatrix}
1 &  1 &  - \\
- &  1 &  -\\
- &  - &  1 \\
1 &  - &  1
\end{bmatrix}.\]
We would then repeat on the columns (over all dimensions).
A small example,
$\begin{bmatrix}
1 & -  & 1 & 1 \\
1 & -  & - & - \\
\end{bmatrix}$, demonstrates this approach is
suboptimal, since the normalization shown is optimal
for $2\times1$ and $1\times2$ blocks but not
optimal for $2\times2$ blocks.

\begin{figure}
{\singlespace
 \begin{algorithmic}
\STATE \textbf{input} a cube $C$ 
   \FORALL {dimensions $i$}
     \FORALL {attribute values $v_1$}
     \FORALL {attribute values $v_2$}
 	\STATE $\begin{array}{lll} w_{v_1,v_2} & \leftarrow
           &\mbox{storage cost of slices $\widehat{C}^i_{v_1}$ and $\widehat{C}^i_{v_2}$ using}\\
           && \mbox{ blocks of shape $\underbrace{1\times\ldots\times1}_{i-1}\times2\times \underbrace{1\times\ldots\times1}_{d-i}$}
	         \end{array}$
     \ENDFOR
   \ENDFOR
   \STATE form graph $G$ with attribute values $v$ as nodes and edge
weights $w$
   \STATE solve the weighted-matching problem over $G$
   \STATE order the attribute values
   so that matched values are listed consecutively
  \ENDFOR
 \end{algorithmic}}
\caption{\label{algo:IM}Iterated Matching (IM) Normalization Algorithm}

 \end{figure}
 
\subsection{One-Dense-Chunk Heuristic: iterated Greedy Sort (GS)}

Earlier work~\cite{kase:compress-tr} discusses 
data-cube normalization under a different
HOLAP model, where only one block may be stored densely, but
the block's size is chosen adaptively.  
Despite model differences,
normalizations that cluster
data into a single large chunk intuitively should be useful with our current
model.  We adapted the most successful heuristic identified in the
earlier work and called the result GS for iterated Greedy 
\glossary{name={GS},description={iterated Greedy Sort heuristic}}
Sort (see Fig.~\ref{algo:GS}).  It can be viewed as a variant
of Frequency Sort that ignores portions of the cube that appear
too sparse.

This algorithm's details are shown in
Fig.~\ref{algo:GS} and sketched briefly next.  
Parameter $\rho_{\textrm{break-even}}$ can be set to the break-even density for
HOLAP storage ($\rho_{\textrm{break-even}}=\frac{1}{\alpha d + 1}=\frac{1}{d/2 + 1}$) (see section \ref{section:block-coded}).
The algorithm
partitions every dimension's values into ``dense'' and ``sparse''
values, based on the current partitioning of all other dimensions'
values.  It proceeds in several phases, where each phase cycles
once through the dimensions, improving the partitioning choices  for that
dimension.  The choices are made greedily within a given phase, although
they may be revised in a later phase.
The algorithm often converges well before 20 phases.

\begin{figure}
 {\singlespace
 \begin{algorithmic}
 \STATE \textbf{input} a cube $C$, break-even density $\rho_{\textrm{break-even}}=\frac{1}{d/2 + 1}$
 \FORALL {dimensions $i$}
   \STATE \COMMENT {$\Delta_i$ records attribute values classified as dense (initially, all)}
   \STATE initialize $\Delta_i$ to contain each attribute value $v$
 \ENDFOR

  \FOR {20 repetitions}
   \FORALL {dimensions $i$}
     \FORALL {attribute values $v$}
     \STATE \COMMENT {current $\Delta$ values mark off a subset of the slice as "dense"}
     \STATE
        $\rho_v \leftarrow $ density of $\widehat{C}^i_v$ within 
           $\Delta_1 \times \Delta_2 \times \ldots \times 
                                       \Delta_{i-1} \times \Delta_{i+1}
                                       \times \ldots$
     \IF {$\rho_v < \rho_{\textrm{break-even}}$ and $v \in \Delta_i$} \STATE remove $v$ from $\Delta_i$
     \ELSIF {$\rho_v \geq \rho_{\textrm{break-even}}$ and $v \not \in \Delta_i$} \STATE add $v$ to $\Delta_i$ \ENDIF
  \ENDFOR   
  \IF {$\Delta_i$ is empty}
     \STATE add $v$ to $\Delta_i$, for an attribute $v$ maximizing $\rho_v$
  \ENDIF
 \ENDFOR 
\ENDFOR
 \STATE Re-normalize $C$ so that each dimension is sorted by its final $\rho$ values

 \end{algorithmic}}
\caption{\label{algo:GS}Greedy Sort (GS) Normalization Algorithm}
 \end{figure}

\begin{figure}

\resizebox{.48\textwidth}{.3\textheight}{\includegraphics{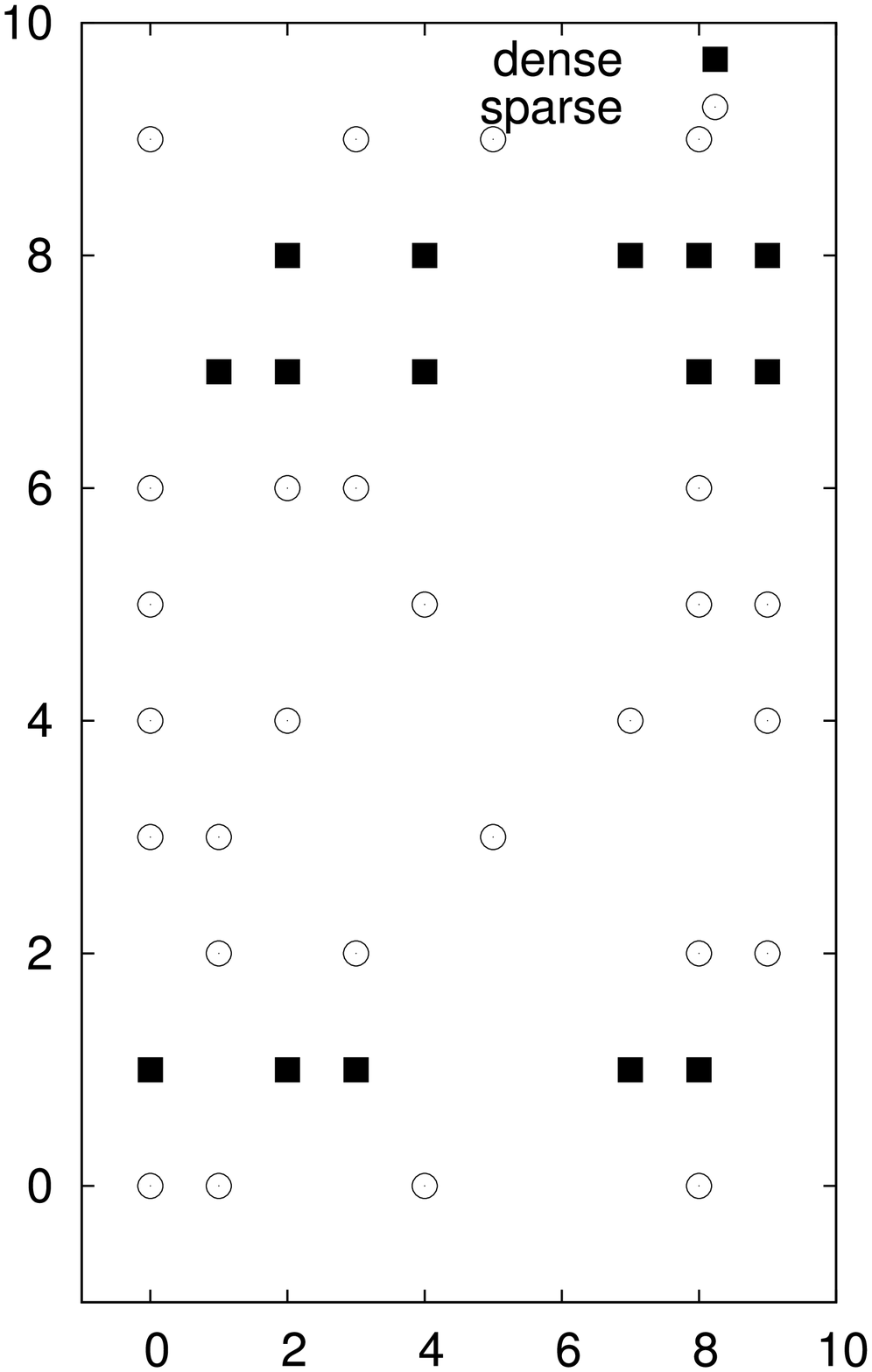}}\hfill
\resizebox{.48\textwidth}{.3\textheight}{\includegraphics{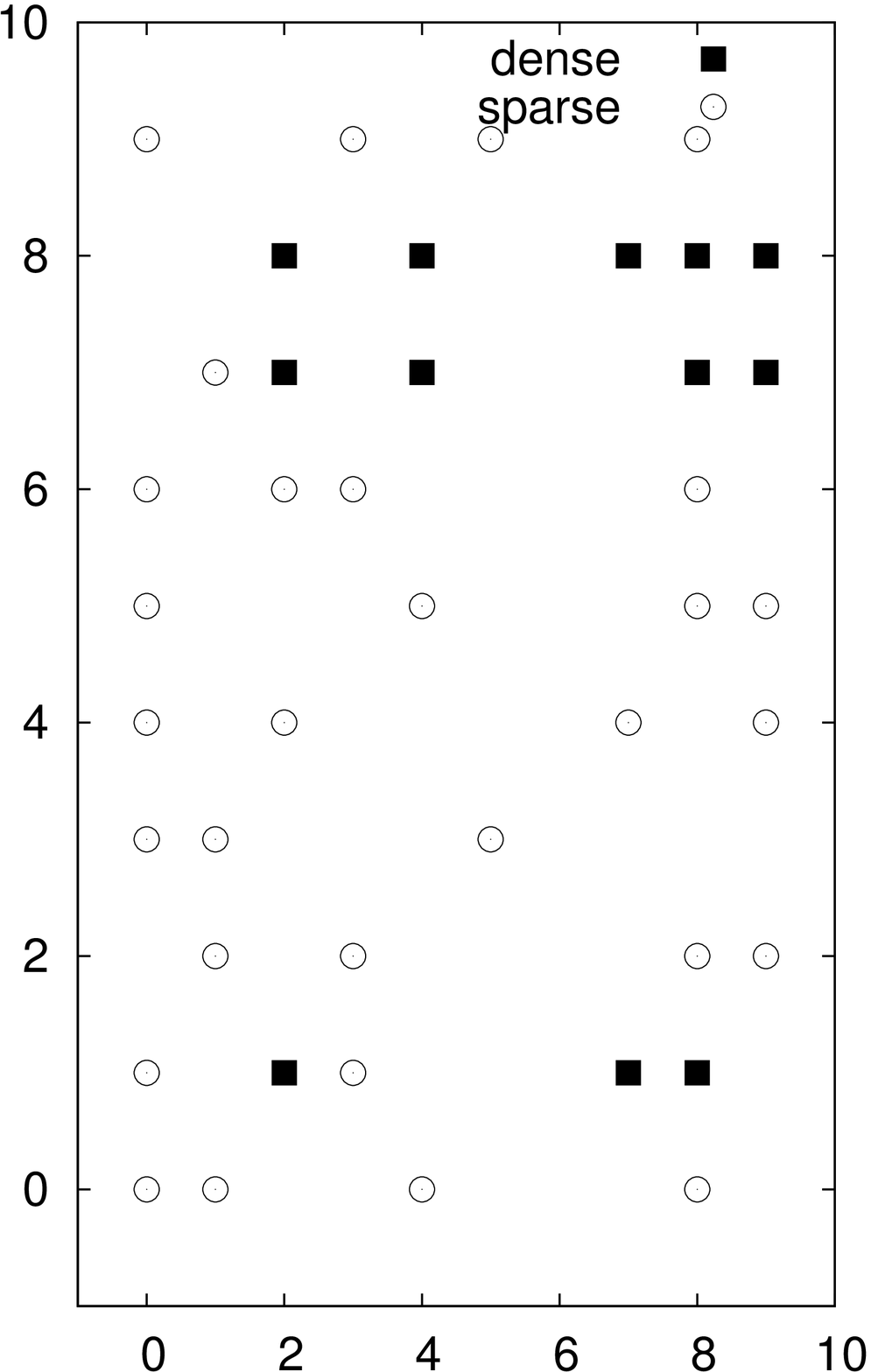}}\\
\resizebox{.48\textwidth}{.3\textheight}{\includegraphics{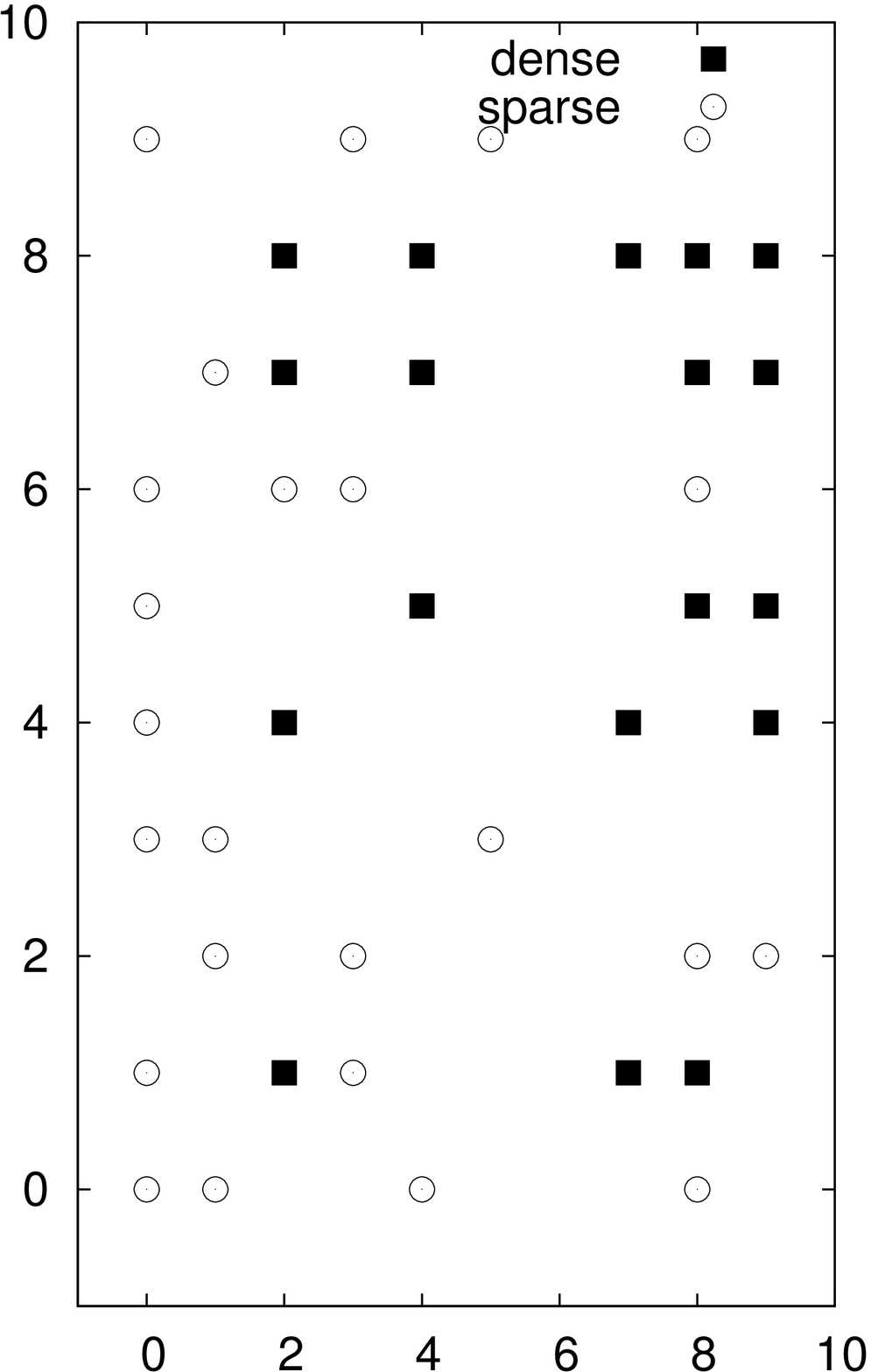}}\hfill
\resizebox{.48\textwidth}{.3\textheight}{\includegraphics{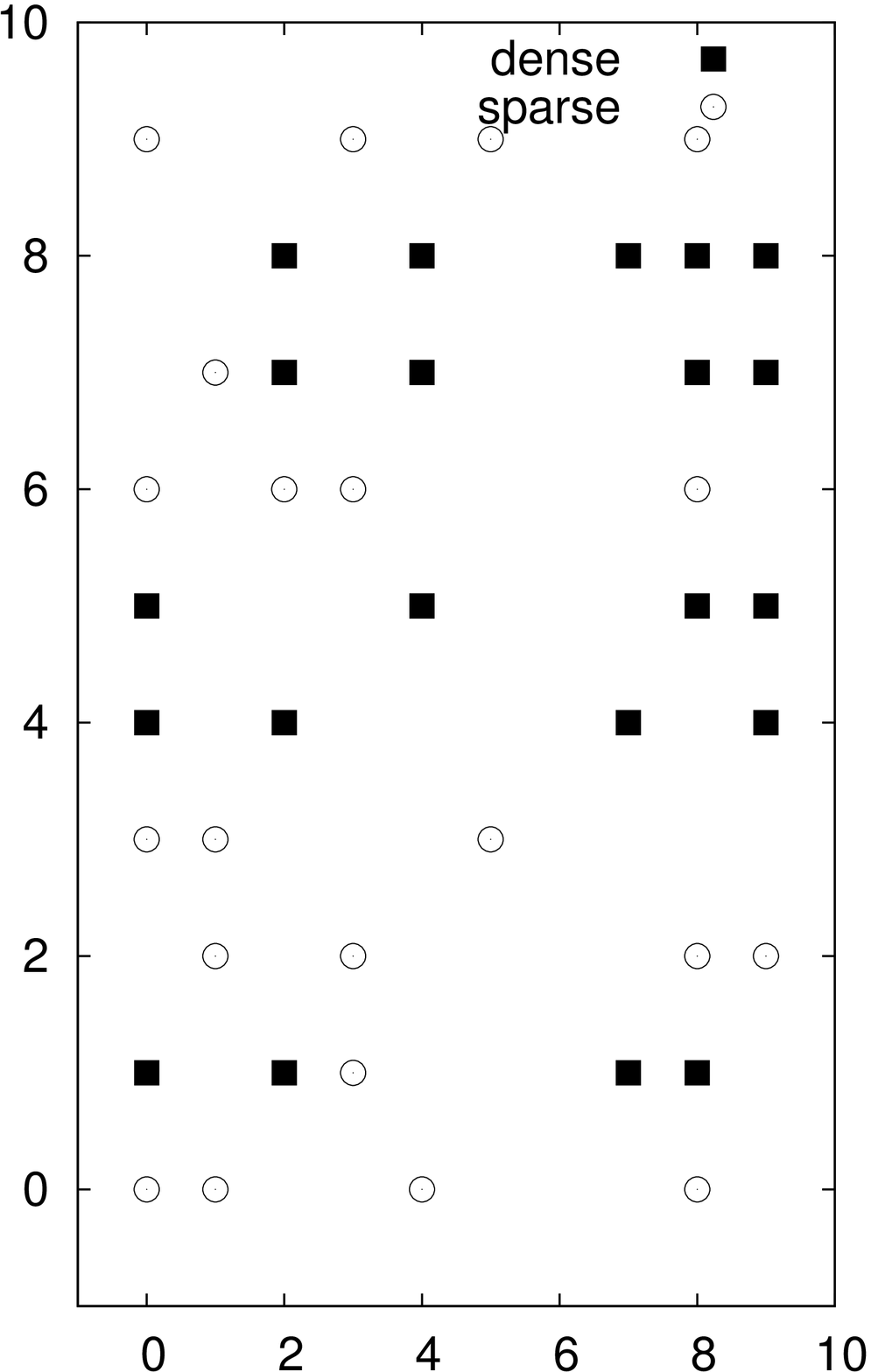}}

\caption{\label{fig:GS-example}GS Example.  Top left: after rows
processed once.  Top right: after columns processed once.  Bottom
left: after rows processed again.  Bottom right: after columns
processed again.}

\end{figure}

Figure~\ref{fig:GS-example} shows GS working over a two-dimensional example
with $\rho_{\textrm{break-even}}=\frac{1}{d/2 + 1}=\frac{1}{2}$. The goal
of GS is to mark a certain number of rows and columns as dense: we would then group
these cells together in the hope of increasing the number of dense blocks.
Set $\Delta_i$ contains
all ``dense'' attribute values for dimension $i$.  Initially,  $\Delta_i$ contains all
attribute values for all dimensions $i$. The initial figure is not shown but
would be similar to the upper left figure, except that all allocated cells would be
marked as dense (dark square). In the upper-left figure, we present 
the result after the rows (dimension $i=1$) have been processed for the first time.
Rows other than 1, 7 and 8 were insufficiently dense and hence removed 
from $\Delta_1$: all allocated cells outside these rows have been marked ``sparse'' (light square).
Then the columns (dimension $i=2$) are processed for the first time, considering
only cells on rows 1, 7 and 8, and the result
is shown in the upper right. Columns 0, 1, 3, 5 and 6 are insufficiently dense
and removed from $\Delta_2$, so a few more allocated cells were marked as sparse (light square). For instance, the density for column
0 is $\frac{1}{3}$ because we are considering only rows 1, 7 and 8.
GS then re-examines the rows (using the new $\Delta_2=\{2, 4, 7, 8 , 9\}$) and reclassifies
rows 4 and 5 as dense, thereby updating $\Delta_1=\{1,4,5,7,8\}$.  Then, when the columns are
re-examined, we find that the density of column 0 has become
$\frac{3}{5}$ and reclassify it as dense ($\Delta_2=\{0,2, 4, 7, 8 , 9\}$).  A few more iterations would be
required before this example converges.  Then we would sort 
rows and columns by decreasing density in the hope that allocated cells would be clustered near cell $(0,0)$.  (If rows 4, 5 and 8 continue
to be 100\% dense, the normalization would put them first.)

\subsection{Summary of heuristics}

Recall that all our heuristics are of the type
``1-dimension-at-a-time'', in that they normalize one dimension at a
time. Greedy Sort (GS) is not orthogonal whereas Iterated Matching
(IM) and Frequency Sort (FS) are: indeed GS revisits the dimensions
several times for different results. FS and GS are block-oblivious
whereas IM assumes 2-regular blocks. The following table is a summary:

\begin{center}\singlespace \footnotesize
\begin{tabular}{|c|c|c|}
 \hline Heuristic & block-oblivious/block-aware &  orthogonal \\ 
  \hline \hline FS & block-oblivious &  true  \\ 
\hline GS & block-oblivious &  false  \\ 
\hline IM & block-aware & true \\
\hline 
\end{tabular}
\end{center}

\section{Experimental Results}
\label{section:experiments}
In describing the experiments, we discuss the data sets used, 
the heuristics tested, and the results observed.

\subsection{Data Sets} 

Recalling that $E(C)$ measures the cost per allocated cell, we define
the \emph{kernel} $\kappa_{m_1,\ldots,m_d}$ as the set of all data cubes
$C$ of given
dimensions such that $E(C)$ is minimal ($E(C)=1$) for some fixed block
dimensions $m_{1},\ldots,m_{d}$. In other words, it is the set of all
data cubes $C$ where all blocks have density $1$ or $0$.
\glossary{name={$\kappa_{m_1,\ldots,m_d}$},description={the kernel: set of cubes such that all blocks of dimensions $m_{1},\ldots,m_{d}$ have density $1$ or $0$}}

Heuristics were tested on a variety of data cubes.  Several 
synthetic $12 \times 12 \times 12 \times 12$ data sets were used, and 100 random data cubes
of each variety were taken.
\begin{itemize}
\item \emph{$\kappa_{2,2,2,2}^{base}$} refers to choosing a cube $C$
uniformly from $\kappa_{2,2,2,2}$ and choosing $\pi$ uniformly from the
set of all normalizations.  Cube $\pi(C)$ provides the test data; a best-possible
normalization will compress $\pi(C)$ by a ratio of $\max(\rho,\frac{1}{3})$,
where $\rho$ is the density of $\pi(C)$. (The expected value of $\rho$ is
50\%.)

\item \emph{$\kappa_{2,2,2,2}^{\mbox{sp}}$} is similar, except that
the random selection from $\kappa_{2,2,2,2}$ is biased towards sparse
cubes.  
\chainsaw{ (Each of the 256 blocks is independently
chosen to be full with probability 10\% and empty with probability
90\%.)
}
The expected density of such cubes is 10\%, and thus the
entire cube will likely be stored sparsely.  The best compression for
such a cube is to $\frac{1}{3}$ of its original cost.

\item \emph{$\kappa_{2,2,2,2}^{\mbox{sp}}$+N} adds noise. For every 
index, there is a 3\% chance
that its status (allocated or not) will be inverted.  Due to the noise, the
cube usually cannot be normalized to a kernel cube, and hence the best
possible compression is probably closer to $\frac{1}{3} + 3\%$.

\item \emph{$\kappa_{4,4,4,4}^{\mbox{sp}}$+N} is similar, except
we choose from $\kappa_{4,4,4,4}$, not $\kappa_{2,2,2,2}$.

\end{itemize}

Besides synthetic data sets, we have experimented with several data
sets used previously~\cite{kase:compress-article}: {\sc Census} (50 6-d
projections of an 18-d data set) and {\sc Forest} (50 3-d projections of
an 11-d data set) from the KDD repository~\cite{KDDRepository},
 and {\sc Weather} (50 5-d projections of an 18-d
data set)~\cite{WeatherSource}\footnote{Projections were selected at random but, to keep test runs
from taking too long, cubes were required to be smaller than about
100MB.}. These data sets were obtained in relational form%
\chainsaw{, as a sequence $\langle t\rangle$ of tuples \ }%
and
their initial normalizations can
be summarized as ``first seen, first when normalized,'' which is
arguably the normalization that minimizes data-cube implementation
effort.  
\chainsaw{
More precisely,
let $\pi$ be the normal relational projection operator; e.g., 
\[\pi_2(\langle (a,b), (c,d), (e,f) \rangle) = \langle b,d,f \rangle.\]
Also let the \emph{rank} $r(v,\langle t \rangle)$ of a value $v$ in a
sequence $\langle t \rangle$ be the number of distinct values that
precede the \emph{first} occurrence of $v$ in $\langle t \rangle$.
The initial normalization for a data set $\langle t \rangle$ permutes
dimension $i$ by $\gamma_i$, where $\gamma^{-1}_i(v) = r(\pi_i(\langle t
\rangle))$.  If the tuples were originally presented in a random
order, commonly occurring values can be expected to be mapped to small
indices: in that sense, the initial normalization resembles
an imperfect Frequency Sort.
This initial normalization has been called ``Order $\mathcal{I}$''
in earlier work~\cite{kase:compress-tr}.
}

\subsection{Results}

The heuristics selected for testing were Frequency Sort (FS),
Iterated Greedy Sort (GS),  and
Iterated Matching (IM).  Except for the
``$\kappa_{4,4,4,4}^{\mbox{sp}}$+N'' data sets, where 4-regular blocks
were used, blocks were 2-regular. IM implicitly
assumes 2-regular blocks. Results are shown in
Table~\ref{heurperf-synth-real}.

\newcommand{\dat}[2] {#1}  

\begin{table}
\caption[Performance of heuristics.]{\label{heurperf-synth-real}Performance of heuristics.
Compression ratios are in percent and are averages.
Each number represents 100 test runs for the synthetic data sets
and 50 test runs for the others.
Each experiment's outcome was the ratio
of the heuristic storage cost to the default normalization's storage cost.
Smaller is better.}
\begin{center}
\begin{singlespace}
\begin{scriptsize}
\begin{tabular}{|l|r|r|r|r|r|r|r|}
\hline
Heuristic                    & \multicolumn{4}{c|}{Synthetic Kernel-Based Data Sets} & \multicolumn{3}{c|}{ ``Real-World'' Data Sets}\\
                             & $\kappa_{2,2,2,2}^{base}$ &
                               $\kappa_{2,2,2,2}^{\mbox{sp}}$ &
                               $\kappa_{2,2,2,2}^{\mbox{sp}}$+N &
                               $\kappa_{4,4,4,4}^{\mbox{sp}}$+N
& {\sc Census} & {\sc Forest} & {\sc Weather}\\ \hline
FS                           &\dat{61.2}{7.9}  &\dat{56.1}{11.6}& \dat{85.9}{7.2} &\dat{\textbf{70.2}}{13.7} & \dat{78.8}{10.6}& \dat{94.5}{3.6} & \dat{88.6}{7.4}  \\
GS                           &\dat{61.2}{7.9}  &\dat{87.4}{7.1} & \dat{86.8}{7.1} &\dat{72.1}{13.8} & \dat{79.3}{11.2}& \dat{94.2}{4.0} & \dat{89.5}{8.5}  \\
IM                           &\dat{\textbf{51.5}}{1.5}  &\dat{\textbf{33.7}}{0.3} & \dat{\textbf{49.4}}{1.3} &\dat{97.5}{4.2} & \dat{\textbf{78.2}}{10.8}& \dat{\textbf{86.2}}{7.2} & \dat{\textbf{85.4}}{7.9}   \\ \hline
Best result (estimated)   &\dat{40}{0}  &\dat{33}{0} & \dat{36}{0} &\dat{36}{0} & --& -- & --   \\ \hline
\end{tabular}
\end{scriptsize}
\end{singlespace}
\end{center}
\end{table}

Looking at the results in Table~\ref{heurperf-synth-real} for
synthetic data sets, 
we see that GS was never better than FS; this is perhaps not
surprising, because the main difference between FS and GS is that
the latter does additional work to ensure allocated cells are within
a single hyperrectangle and that cells outside this hyperrectangle
are discounted. 

 Comparing
the $\kappa_{2,2,2,2}^{\mbox{sp}}$ and $\kappa_{2,2,2,2}^{\mbox{sp}}$+N  
columns, it is apparent that noise hurt all
heuristics, particularly the slice-sorting ones (FS and GS).
However, FS and GS performed better on larger blocks ($\kappa_{4,4,4,4}^{\mbox{sp}}$+N)
than on smaller ones ($\kappa_{2,2,2,2}^{\mbox{sp}}$+N) whereas 
IM did worse on larger blocks. We explain this improved performance 
for slice-sorting normalizations (FS and GS) as follows: $\#C^i_v$ is a multiple
of $4^3$ under $\kappa_{4,4,4,4}$ but a multiple of $2^3$ under $\kappa_{2,2,2,2}$.
Thus, $\kappa_{2,2,2,2}$ is more susceptible to noise than  $\kappa_{4,4,4,4}$
under FS because the values $\#C^i_v$ are less separated. IM did worse on
larger blocks because it was designed for 2-regular blocks.

Table~\ref{heurperf-synth-real} also contains results for
``real-world'' data, and the relative performance of the various
heuristics depended heavily on the nature of the data set used.
\verbose{Part of this is due to the nature of the data sets: for} For
instance, {\sc Forest} contains many measurements of physical
characteristics of geographic areas, and
\verbose{there is} significant correlation between
\verbose{these} characteristics \verbose{that tended to} penalized FS.

\subsubsection{Utility of the Independence Sum}
\begin{figure} 
\begin{center}{\resizebox{13cm}{!}{\includegraphics{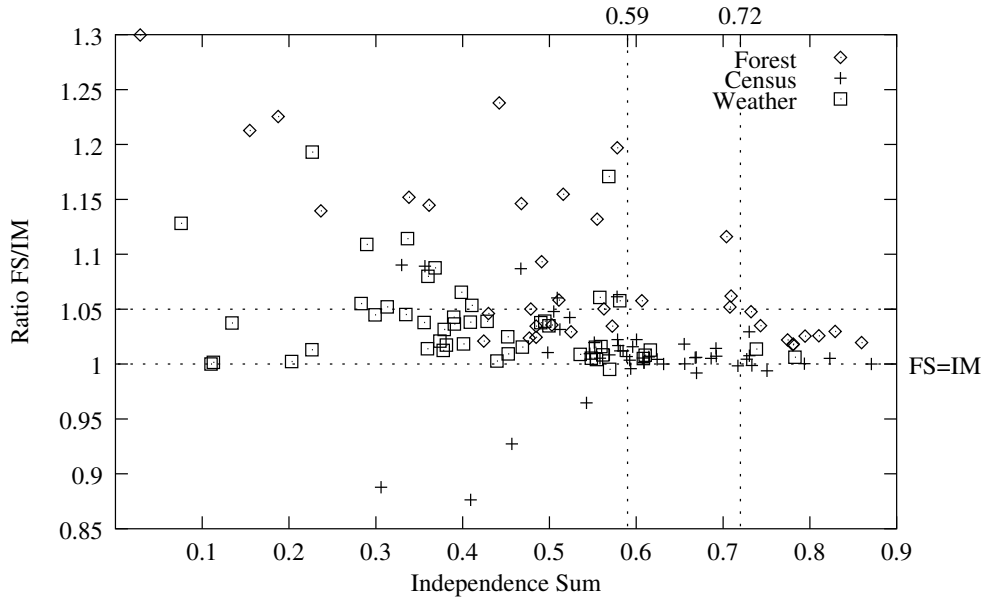}}}\end{center}
\caption[Solution-size ratios of FS and IM as a function of
Independence Sum.]{\label{indepsum-fig}Solution-size ratios of FS and IM as a function of
Independence Sum.  When the
ratio is above 1.0, FS is suboptimal; when it is less than 1.0, IM
is suboptimal. We see that as the Independence Sum approached
1.0, FS matched IM's performance.}
\end{figure}

Despite the differences between data sets, the Independence Sum (from
Section~\ref{section:slicesort}) seems to be useful.  In
Figure~\ref{indepsum-fig} we plot the ratio $\frac{\mbox{size using
FS}}{\mbox{size using IM}}$ against the Independence Sum.  
When the Independence Sum exceeded 0.72, the ratio was always near 1
(within 5\%); thus, there is no need to use the more computationally
expensive IM heuristic.  {\sc Weather} had few cubes with Independence
Sum over 0.6, but these had ratios near 1.0.  For {\sc Census},
having an Independence Sum over 0.6 seemed to guarantee good relative
performance for FS.  On {\sc Forest}, however, FS showed poorer
performance until the Independence Sum became larger ($\approx 0.72$).

\subsubsection{Density and Compressibility}

\begin{figure} 
\begin{center}
\resizebox{12cm}{!}{\includegraphics{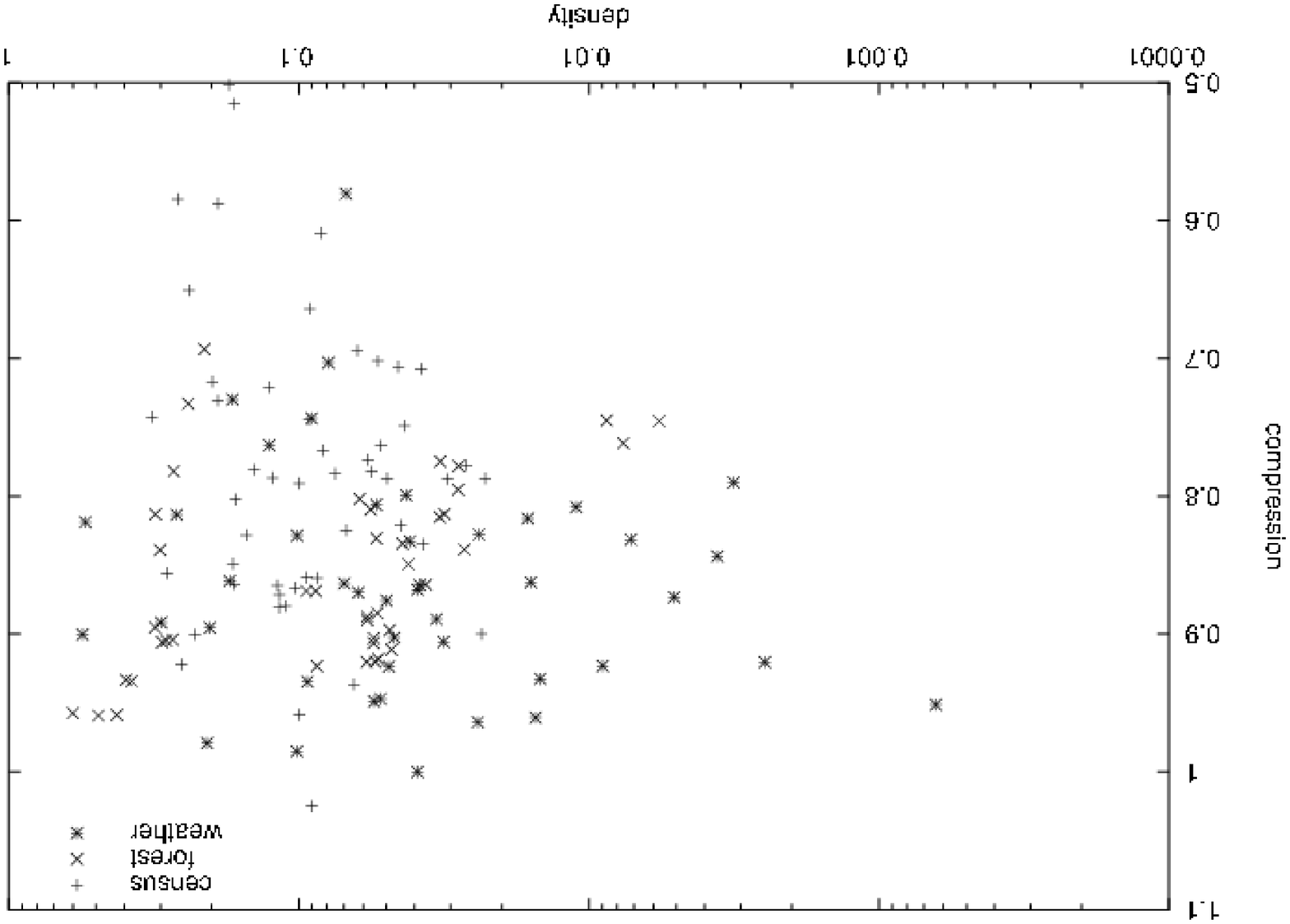}  }
\resizebox{12cm}{!}{\includegraphics{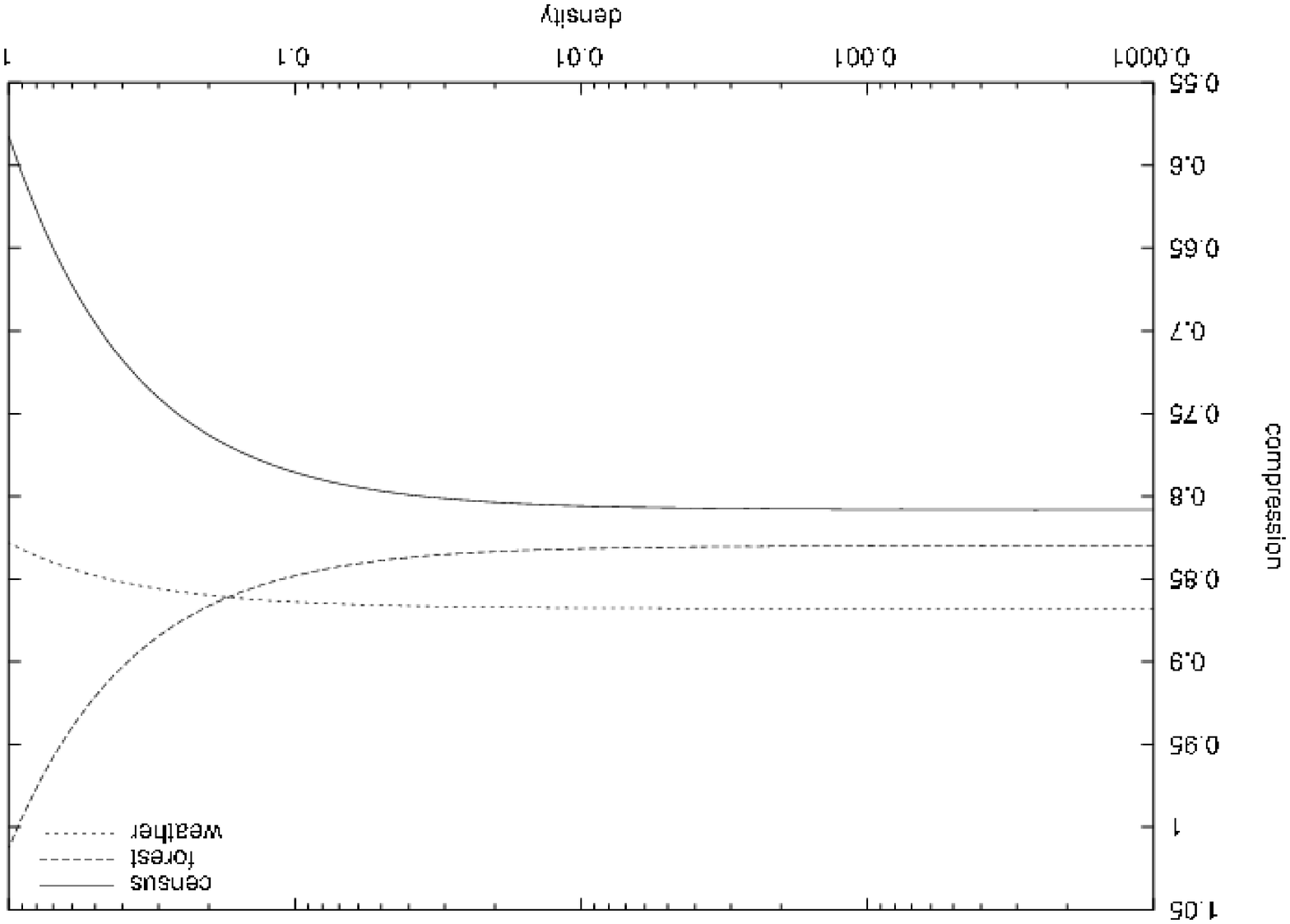}}
\end{center}
\caption[Compression ratios achieved with IM versus density
for 50 test runs on three data sets.]{\label{fig-density}
Compression ratios achieved with IM versus density
for 50 test runs on three data sets. The bottom plot shows linear
regression on a logarithmic scale: both \textsc{Census} and 
\textsc{Weather} showed
a tendency to better compression with higher density.
}
\end{figure}

The results of Table~\ref{heurperf-synth-real} are averages over cubes
of different densities. Intuitively, for very sparse cubes (density
near 0) or for very dense cubes (density near 100\%), we would expect
attribute-value reordering to have a small effect on compressibility:
if all blocks are either all dense or all sparse, then attribute
reordering does not affect storage efficiency. We take the source data
from Table~\ref{heurperf-synth-real} regarding Iterated Matching (IM)
and we plot the compression ratios versus the density of the cubes
(see Fig.~\ref{fig-density}). \verbose{Our results indicate that  2 out of 3}
Two of three
data sets showed some compression-ratio improvements when the
density is increased, but the results are not conclusive. An
extensive study of a related problem is described 
elsewhere~\cite{kase:compress-tr}.

\subsubsection{Comparison with Pure ROLAP Coding}

To place the efficiency gains from normalization into
context, we calculated (for each of the 50 {\sc Census} cubes)
$c_{\rm default}$, the HOLAP storage cost using  2-regular blocks
and the default normalization.
We also calculated $c_{\rm ROLAP}$, the ROLAP cost, for each
cube. The average of the 50 ratios $\frac {c_{\rm default}}
{c_{\rm ROLAP}}$ was 0.69 with a standard deviation of 0.14.
In other words, block-coding was 31\% more efficient than
ROLAP. On the other hand, we have shown that normalization
brought gains of about 19\% over the default normalization and
the storage ratio itself was brought from 0.69 to 0.56
in going from simple block coding to block coding together
with optimized normalization.
{\sc Forest} and {\sc Weather} were similar, and their
respective average ratios $\frac {c_{\rm default}}
{c_{\rm ROLAP}}$ were 0.69 and 0.81.  Their respective
normalization gains were about 14\% and 12\%, resulting
in overall storage ratios of about 0.60 and 0.71, respectively.
\verbose{Another way to look at the result is that the overall
storage ratio decreased from 0.69 to 0.56, from 0.69 to 0.60,
and from 0.81 to 0.71, respectively: a decrease of around 0.1
in each case.}

\section{Conclusion}
\label{section:conclusion}

In this paper, we have given several theoretical results relating
to cube normalization.  Because even simple special cases of the
problem are NP-hard, heuristics are needed.  However, an optimal
normalization can be computed when $1\times2$ blocks are used, and
this forms the basis of the IM heuristic, which seemed most efficient
in experiments.  Nevertheless, a Frequency Sort algorithm is much faster,
and another of the paper's theoretical conclusions was that this
algorithm becomes increasingly optimal as the Independence Sum of
the cube increases: if dimensions are nearly statistically independent,
it is sufficient to sort the attribute values for each dimension separately.
Unfortunately, our theorem did not provide a
very tight bound on suboptimality.  Nevertheless, we determined
experimentally that an Independence Sum greater than 0.72 always meant
that Frequency Sort produced good results.

As future work, we will seek
tighter theoretical bounds and more effective heuristics for the cases
when the Independence Sum is small. We are \verbose{also in the process of}
implementing the proposed architecture \verbose{for further validation and research}
by combining an embedded relational database with a C++ layer.
We will verify our claim that
a more efficient normalization leads to faster queries.

\section*{Acknowledgements}

The first author was supported in part by NSERC grant 155967
and the second author was supported in part by NSERC grant 261437.
The second author was \verbose{a research officer} at the National Research Council
of Canada when he began this work.

\typeout{Typesetters, beware...the glossary following references page numbers.  It assumes page 1 is the first page of this article.}

\newpage
{\singlespace \footnotesize \printglossary}

{\singlespace \footnotesize \bibliographystyle{elsart-num}
\addcontentsline{toc}{section}{\refname}\bibliography{is2004}
}
\end{document}